\documentclass[authoryear,12pt]{elsarticle}

\usepackage{graphicx}
\usepackage{wrapfig}
\usepackage{lscape}
\usepackage{rotating}
\usepackage{epstopdf}
\usepackage{blindtext, rotating}
\usepackage{caption}

\usepackage{subcaption}

\usepackage{amssymb,amsmath}
\usepackage{exscale}
\usepackage{makeidx,shortvrb,latexsym}
\usepackage{epstopdf}
\usepackage{tabularx, booktabs, multirow}
\usepackage[]{hyperref}
\newcommand{\rd}{\mathrm{d}}
\usepackage{color}
\journal{Journal of the Mechanics and Physics of Solids}

\begin{document}

\begin{frontmatter}

\title{Simulation of the Hall-Petch effect in FCC polycrystals by means of strain gradient crystal plasticity and FFT homogenization}

\author[myaddress]{S.~Haouala }
\author[myaddress,mysecondaryaddress]{S.~Lucarini}
\author[myaddress,mysecondaryaddress]{J.~LLorca\corref{cor}}
\ead{javier.llorca@imdea.org}
\author[myaddress,mysecondaryaddress]{J.~Segurado\corref{cor}}
\ead{javier.segurado@imdea.org}
\address[myaddress]{IMDEA Materials Institute, C/ Eric Kandel 2, 28906 - Getafe, Madrid, Spain}
\address[mysecondaryaddress]{Department of Materials Science, Polytechnic University of Madrid, E. T. S. de Ingenieros de Caminos, 28040 - Madrid, Spain}
\cortext[cor]{Corresponding authors.}

\begin{abstract}

The influence of grain size on the flow stress of various FCC polycrystals (Cu, Al, Ag and Ni) has been analyzed by means of computational homogenization of a representative volume element of the microstructure using a FFT approach in combination with a strain gradient crystal plasticity model. The density of geometrically necessary dislocations resulting from the incompatibility of plastic deformation among different crystals was obtained from the Nye tensor, which was efficiently obtained from the curl operation in the Fourier space. The simulation results were in good agreement with the experimental data  for Cu, Al, Ag and Ni polycrystals for grain sizes $>$ 20 $\mu$m and strains $ <$ 5\% and provided a physical explanation for the higher strengthening provided by grain boundaries in Al and Ni, as compared with Cu and Ag. The investigation demonstrates how the combination of FFT with strain gradient crystal plasticity can be used to include effect of grain boundaries in the mechanical behavior of polycrystals using realistic representative volume elements of the microstructure.

\end{abstract}


\end{frontmatter}

\section{Introduction}

Computational homogenization of polycrystals has become a very powerful tool to simulate the mechanical behavior of polycrystalline aggregates \citep{REH10, SLL18}. Within this framework, the effective properties of the polycrystal are determined by means of the full-field solution of a boundary value problem of a Representative Volume Element (RVE) of the microstructure under homogeneous boundary conditions. It is well established that the success of the approach relies in two factors: the accurate representation of the microstructure in the digital model (including the appropriate information about the grain size, shape, texture, etc.) and the constitutive description of the single crystal behavior by means of crystal plasticity, which nicely captures the deformation kinematics induced by dislocation slip. Nevertheless, the effect of grain boundaries is not included in most cases and the development of sound strategies to account for their influence on the deformation mechanisms and flow stress is an important goal of computational homogenization of polycrystals.

Grain boundaries are often barriers to dislocation slip (but not always) and modify the deformation mechanisms in two ways. First, they promote heterogeneous deformation of the single crystals within the polycrystal, so the activity of the slip systems in each grain differs significantly within the grain as a function of the distance to the boundary or of the behavior of the nearest boundary \citep{DRC00, HNV18, BAP19}. Second, the opaque grain boundaries lead to the formation of dislocation pile-ups, increasing the critical resolver shear stress to move dislocations and, thus, leading to the well-known Hall-Petch behavior \citep{Hall1951-1,Petch1953-1}. Different theoretical models have been developed in the past to quantify the strengthening of grain boundaries \citep{A70, K70, H72, LBD16} but it is widely accepted today that this phenomenon is a manifestation of the general size effect found in plasticity which depends on many factors, such as the elastic anisotropy of the crystal, the range of grain sizes examined, the texture, the number of slip systems, the initial dislocation density, the presence of other obstacles to dislocation motion, etc. \citep{BDN08, FBM01}. All these features can be accurately taken into account through the computational homogenization of poycrystals that becomes a very useful tool to make quantitative predictions of the influence of the grain size on the strength of polycrystals.

Homogenization models of polycrystals based on classical plasticity cannot capture the grain size effect because the constitutive equation does not involve an intrinsic materials length scale. Nevertheless, this limitation can be overcome by introducing a length associated with the plastic strain gradients. The effect of the plastic strain gradients in the mechanical behavior  was first introduced by \cite{AIFANTIS1987211} for homogenized plasticity and further extended to crystal plasticity to account for size effects in the deformation of single crystals or bicrystals \citep{ACHARYA1995,SHU1999297}. These seminal papers defined two different approaches for strain gradient crystal plasticity (SGCP). In the first one, denominated  \emph{lower-order} SGCP, the gradients of the plastic slip -- accounted as an internal variable -- are included in the hardening moduli. This framework preserved the thermodynamic consistency  \citep{Acharya2003-1} and allows the use of the classical mathematical framework of boundary value problems and constitutive equations with internal variables \citep{Dai1997-1,Busso2000-1,HAN20051188,HAN20051204,Ma20062169,DUNNE20071061}. In the second approach, \emph{higher order} SGCP, the internal variables from which gradients are extracted are chosen as kinematic variables. This implies the resolution of a boundary value problem of a system of coupled partial differential equations, in which both the displacements and some fields related to the plastic strain are solved. A consequence of this choice is the presence of stresses conjugated to these new kinematic variables as well as the corresponding "higher order" boundary conditions \citep{GURTIN20025,GURTIN2008702, BARDELLA2006128, BSP13, NIORDSON201431}.

Lower-order  \citep{AB00, CBA05, BBG07, BER10} and higher-order SGCP models \citep{BER10},  have been used within the framework of computational homogenization to analyze the effect of grain size in polycrystals using the finite element method. Although both approaches predict a size effect, a detailed comparison with abundant experimental data in the literature is lacking because most of the simulations were limited to small representative volume elements containing a few dozens of grains because of computational limitations. In the case of lower-order approaches, the computational difficulties are associated with obtaining the plastic strain gradients from the finite element framework. They are computed within each element from the derivative of a field constructed from the plastic strain at the integration points \citep{Busso2000-1,DUNNE20071061}, leading to a non-local approach that implies a discontinuity of the gradient field from element to element and may lead to convergence problems. Higher-order approaches are computationally much more expensive because the number of degrees of freedom per node in three dimensions increases from 3 to 12 when the plastic deformation gradients are added to the displacements as kinematic variables \cite{GURTIN20025}. Moreover, the use of higher order theories within a polycrystalline framework present some theoretical difficulties since grain boundaries are not external boundaries and variables associated with the slip of a particular system are only well defined inside each grain. Although some of these difficulties are theoretically solved  by the introduction of grain boundary Burgers tensors and their corresponding constitutive laws \cite{Gurtin2005,Gurtin2008}, the numerical implementation of the resulting models is even more challenging. 

Some of the computational limitations indicated above can be overcome with the use of the fast Fourier transform (FFT) to solve the boundary value problem of the RVE of the polycrystal.  The original approach for linear elastic materials solved the equilibrium equation in a heterogeneous medium using a reference material and the Green's functions method \citep{MS98}. The application of the FFT  to polycrystals was pioneered by \cite{L01} and, since then,  crystal plasticity has been used in combination with FFT to determine the effective properties of very large polycrystalline RVEs \citep{LLP04,Eisenlohr2013,ROVINELLI2019, ROTERS2019420} taking advantage of the outstanding computational efficiency of this method. In fact, \cite{LN16} implemented higher order SGCP model in small strains a using a FFT solver and discrete Fourier derivatives \citep{BERBENNI2014}. The model was based on Gurtin's approach \citep{GURTIN20025} and phenomenological crystal plasticity and was used to analyze the mechanical behavior of a polycrystal containing 27 grains, reproducing the grain size effect for a micro-clamped (inpenetrable grain boundaries) condition.

In this paper, an alternative homogenization framework for strain gradient crystal plasticity based on the Galerkin FFT approach \citep{Vondrejc2014} is developed. The approach relies on lower-order SGCP combined with a physically-based crystal plasticity model under finite deformations, allowing the simulation of RVEs containing several hundreds of grains with different average grain sizes. The model is used to analyze the effect of grain size on a number of FCC polycrystals (Cu, Al, Ni and Ag) and all the CP model parameters have a clear physical meaning and were identified for each FCC crystal from either experiments or dislocation dynamics. The results of the simulations were compared with the experimental data in the literature for the Hall-Petch law and provided a detailed understanding of the role played by plastic strain gradients on the strengthening mechanisms in FCC metallic polycrystals.

\section{Computational homogenization framework}
\subsection{Strain gradient crystal plasticity model}\label{sec1}

A physically-based, lower-order SGCP model has been developed to represent the crystal behavior within the FFT-based computational homogenization framework. The model is formulated in finite strains, and the equilibrium is solved in the reference configuration, being the deformation gradient the kinematic variable while the equilibrium is established through the first Piola-Kirchoff stress. The deformation gradient, $\mathbf{F}$, is decomposed into the elastic, $\mathbf{F}^\mathrm{e}$, and plastic components, $\mathbf{F}^\mathrm{p}$,

\begin{equation}
\label{eq_(1):FDecomposition}
\mathbf{F}=\mathbf{F}^\mathrm{e}\mathbf{F}^\mathrm{p}
\end{equation}

\noindent and the plastic velocity gradient  in the intermediate configuration, $\mathbf{L}^\mathrm{p}$, is given by the sum of the shear rates $\dot{\gamma}^\alpha$ on all the slip systems $\alpha$ according to

\begin{equation}
\label{eq_(2):Lp}
\mathbf{L}^\mathrm{p} =\dot{\mathbf{F}}^\mathrm{p}\mathbf{F}^{\mathrm{p}^{-1}}=\sum_\alpha \dot{\gamma}^\alpha \mathbf{s}^\alpha \otimes \mathbf{m}^\alpha
\end{equation}

\noindent  where $\mathbf{s}^\alpha$ and $\mathbf{m}^\alpha$ stand for the unit vectors in the slip direction and the normal to the slip plane, respectively, in the reference configuration.

The elastic deformation,	$\mathbf{E}^\mathrm{e}$, is characterized by the Green-Lagrange deformation tensor

\begin{equation}
\label{eq_(3):Green-Lagrange}
\mathbf{E}^\mathrm{e}=\frac{1}{2}\left( {\mathbf{F}^\mathrm{e}}^T {\mathbf{F}^\mathrm{e}}-\mathbf{I}\right)
\end{equation}

\noindent  where $\mathbf{I}$ is the second order identity tensor. The symmetric second Piola-Kirchhoff stress tensor in the intermediate configuration, $\mathbf{S}$, is related with the Green-Lagrange strain tensor according to

\begin{equation}
\label{eq:hook}
\mathbf{S}=\mathbb{C}:\mathbf{E}^\mathrm{e}
\end{equation}

\noindent  where $\mathbb{C}$ is for the fourth order elastic stiffness tensor of the single crystal. 

The driving force for the plastic slip is the resolved shear stress $\tau^\alpha$ on the slip plane $\alpha$ and it is obtained as the projection of the second Piola-Kirchhoff stress on the slip system according to

\begin{equation}
\label{eq_(5):resolved_shear stress}
\tau^\alpha=\mathbf{S}:(\mathbf{s}^\alpha \otimes \mathbf{m}^\alpha).
\end{equation}

Finally, the plastic slip rate follows a power-law expression,
\begin{equation}
\dot{\gamma}^{\alpha}=\dot{\gamma}_0\left(\frac{|\tau^{\alpha}|}{\tau^{\alpha}_c}\right)^m \mathrm{sgn}(\tau^{\alpha})
\end{equation}

\noindent where the critical resolved shear stress, $\tau^{\alpha}_c$,  is given by a physically-based model including the effect of slip gradients. In particular,  $\tau^{\alpha}_c$,  is given by the Taylor model generalized by \cite{FBZ80} to account for the anisotropy of the interactions between different slip systems according to

\begin{equation}
\tau^{\alpha}_c = \mu b\sqrt{\sum_{\beta}q^{\alpha \beta} \rho^{\beta}}.
\label{eq:Taylor_hard}
\end{equation}

The values $q^{\alpha \beta} $ are dimensionless coefficients that represent the average strength of the interactions between dislocations in pairs of slip systems and that have been computed by means of dislocation dynamics simulations of FCC polycrystals \citep{interaction-matrix-coefficients}.

The dislocation density for each slip system $\rho^\alpha = \rho^\alpha_{SSD} + \rho^\alpha_{GND}$ is split in two terms that encompass the Statistically Stored Dislocations (SSD), $\rho^\alpha_{SSD}$, and the Geometrical Necessary Dislocations (GND),  $\rho^\alpha_{GND}$. The plastic strain gradient enters in the model through a measure of the GND density. In this model, the intrinsic characteristic length relating slip gradients with hardening is the Burgers vector since the hardening is computed using the Taylor model (eq. \ref{eq:Taylor_hard}) which includes the GND density. This approach is therefore similar to the mechanism-based model proposed in \cite{HAN20051188}, in which the characteristic length is also the Burgers vector, although our model accounts for the particular GND density in each slip system instead of using an effective value for all slip systems.

The evolution of the  density of SSD is given by the balance between the rate of accumulation and annihilation of dislocations according to the Kocks-Mecking law \citep{KM81} 

\begin{equation}
\label{eq:kocks}
 \dot{\rho}^{\alpha}_{SSD}=\frac{1}{b} \left(\frac{1}{\ell^{\alpha}}- 2y_c \rho^{\alpha}_{SSD} \right)|\dot{\gamma}^{\alpha} |
  \end{equation}
  
\noindent where $\ell^{\alpha}$ stands for average mean-free path of dislocations in system $\alpha$ and $y_c$ is the effective annihilation distance. The value of the mean free path in the model is given by

\begin{equation}
\label{eq:meanfree}
\ell^{\alpha}=\frac{K}{\sqrt{\sum_{\beta \neq \alpha} \rho_{SSD}^\beta+\rho_{GND}^\beta} }
\end{equation}

\noindent where  $K$ is the similitude coefficient \citep{SK11}.  It should be noted that the dislocation mean free path is affected by both SSDs and GNDs and  decreases  in the areas with higher plastic distortion. 

The density of GNDs is driven by the kinematic constrain to maintain the continuity in the crystal in the presence of gradients in the plastic strain field. GNDs were introduced by \cite{Nye1953-1} who proposed a tensorial magnitude, the Nye tensor, $\mathbf{G}$, as a generalized measure of lattice curvature. The definition of the Nye tensor in terms of GND densities, introduced by \cite{Nye1953-1}, was generalized by \cite{ARSENLIS19991597} as

\begin{equation}
\label{defA}
G_{ij} = \sum_a \rho_{GND}^a b_i^a  t_j^a
\end{equation}

\noindent where $a$ stands for a straight dislocation segment of length $l^a$ parallel to $\mathbf{t}^a$ with Burgers vector $\mathbf{b}^a$,  and $\rho_{GND}^a$ stands for the length of the segment per unit volume. If the number of pure edge and screw segments $a$ in the lattice is $N_{sl}$, this expression can be written in a matrix form

\begin{equation}
\label{defA2}
\bar{G} = \mathbf{A}(\mathbf{b}^a,\mathbf{t}^a)\bar{\rho}_{GND}
\end{equation}

\noindent where $\bar{G}$ is the vector representation of the Nye tensor, $\bar{\rho}_{GND}$ is a vector of dimension $N_{sl}$ containing the GND density on each system $a$ and $\mathbf{A}$ is a matrix of dimensions $9 \times N_{sl}$.

Nye's tensor accounts for the lattice curvature and, therefore, can be expressed as function of the plastic slip gradients. 
Different approaches can be found  in the literature to extend the definition of the Nye tensor in small strains \citep{Nye1953-1} to finite deformations, see \cite{CERMELLI20011539}.  In this work, the Nye tensor is taken as

\begin{equation}
\label{Nye_finite_strain}
\mathbf{G}=\nabla \times  \mathbf{F}^\mathrm{p}.
\end{equation} 

The relation between the Nye tensor defined in terms of GND densities or expressed as function of the plastic strain gradients (eqs. \eqref{defA} and \eqref{Nye_finite_strain}) is general for any crystal lattice. When the expression (\ref{defA}) is particularized for a FCC lattice, multiple combinations of pure edge and screw dislocation segments $a$ are compatible with the same geometric properties, and the nine independent components of the dislocation tensor do not determine uniquely the GND density of each segment $a$. \cite{ARSENLIS19991597} showed that the crystallographic dislocation density derived from the constraints must be minimum for a given value of the Nye tensor in the absence of any other extra information. Two constrained minimization procedures were proposed by \cite{ARSENLIS19991597}, and the minimization of the sum of the squares of GND densities has been chosen here because a close-form for the minimum can be derived. 

In the case of a FCC lattice, the number of independent dislocation segments is $N_{sl}$=18, corresponding to 12 edge dislocations with $\mathbf{t}$ =$<$112$>$ and $\mathbf{b}$=$<$110$>$ and  6 screw segments with $\mathbf{t}=\mathbf{b}$=$<$110$>$. The GND density obtained by $L_2$ minimization is given by

\begin{equation}
\label{mingnd}
\bar{\rho}_{GND}=\mathrm{argmin}_{\bar{\rho}} \{ \bar{\rho}^T\bar{\rho} + \lambda(\mathbf{A}\bar{\rho}-\bar{G}) \}= \left( \mathbf{A}^T\mathbf{A}\right)^{-1} \mathbf{A}^T \bar{G} 
\end{equation}

\noindent where $ \lambda$ are the Lagrange multipliers introduced to constrain eq. \eqref{defA2}. 

\subsection{FFT polycrystalline homogenization}

\begin{figure}[t]
\begin{center}
\includegraphics[width=100mm]{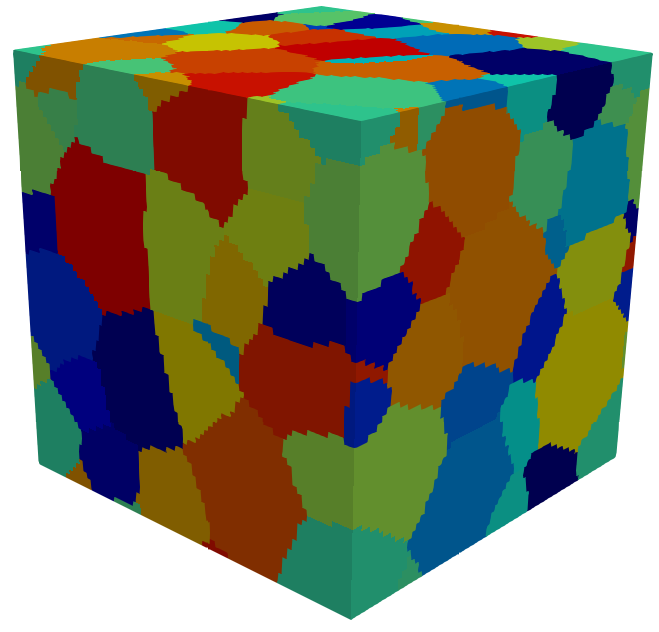}
\end{center}
\caption{RVE containing 100 crystals discretized with 75 x 75 x 75 voxels.}\label{RVE}
\end{figure}

A polycrystalline homogenization framework based on a FFT solver was used to obtain the mechanical response of the polycrystalline RVEs. To this end, periodic RVEs containing 100 and 200 grains were generated using Dream3D \cite{DREAM3D}. The grains within the RVE were equiaxed and the grain sizes followed a lognormal distribution with an average grain size $\overline{D}_g$, standard deviation of 0.2$\overline{D}_g$ and random texture. The RVEs were discretized in an array of $N^3$ equispaced voxels (Figure \ref{RVE}), with $N$ = 50, 75 and 100 to check that the results were independent of the grid size and this goal was achieved with $N$  = 100 and 200 crystals in the RVE. Reducing the grid to $N$ = 75 (to reduce the computational effort) and using only 100 crystals in the RVE (to maintain approximately the same number of voxels per grain) led to very similar results (reduction in the flow stress of $\approx$ 4\% in comparison with $N$ = 100 and 200 crystals). Moreover, simulations with different grain realizations with $N$ = 75 and 100 grains in the polycrystal presented very limited scatter ($\pm$ 2\%), indicating that the effective properties of the polcrystals could be obtained with RVEs containing 100 grains. Thus, the simulations presented below were carried out with $N$ = 75 and 100 grains in the RVE. 

The mechanical response of the RVE under a given macroscopic deformation history is obtained using the code FFTMAD \citep{LS2019a,LS2019b}. The model developed is based on the Galerkin FFT approach \citep{Vondrejc2014,Geers2016,Geers2017} which has been extended to account for the effect of plastic strain gradients. The FFT homogenization approach aims at obtaining the microscopic periodic deformation gradient field in the RVE that fulfils the equilibrium and compatibility conditions. Equilibrium is expressed in the reference configuration using the first Piola-Kirchoff stress, given by
\begin{equation}
\label{eq:Piola}
\mathbf{P(F},\boldsymbol{\alpha})=\mathbf{F}^\mathrm{e}\mathbf{S}{\mathbf{F}^\mathrm{e}}^T\mathbf{F}^{-T}
\end{equation}

\noindent where $\mathbf{F}^\mathrm{e}$ and $\mathbf{S}$ are given for each material point by the constitutive model described in the previous section. $\boldsymbol{\alpha}$ is a vector containing the current value of the internal variables, namely the elastic deformation gradient, the SSD densities in each slip system, and the dislocation Nye tensor $\mathbf{G}$, eq. \eqref{Nye_finite_strain}. The dislocation Nye tensor is computed at a given time for all the points in the RVE from the curl operation in the Fourier space.

The macroscopic loading history is set by defining 9 independent components of the macroscopic deformation gradient or the first Piola-Kirchoff stress history, $\overline{F}_{ij}(t)$ and $\overline{P}_{IJ}(t)$, following the mixed control approach developed by \cite{LS2019b}. The solution of the homogenization problem at a given time $t$ is given by the local value of the periodic deformation gradient $\mathbf{F}(\mathbf{x})$ that fulfills the linear momentum balance in a weak form

\begin{equation}\label{weakarbi}
\left\{ \begin{array}{cl}
\int_{\Omega} \mathbb{G}^* \ast \boldsymbol{\zeta}(\mathbf{x}): \mathbf{P}\left(\mathbf{F}(\mathbf{x}),\boldsymbol{\alpha})\right)\mathrm{d}\Omega & =0\\
 \left<P\right>_{IJ} &=\overline{P}_{IJ}(t) \\
 \left<F\right>_{ij} & =\overline{F}_{ij}(t)
\end{array} \right. 
\end{equation}

\noindent where $<>$ states for the volumetric average, $i,j \cap I,J = \emptyset$ and $\boldsymbol{\zeta}(\mathbf{x})$ represents the test functions, second order tensors convoluted by a projector operator $\mathbb{G}^*$, to enforce their compatibility.

The Galerkin approach of the equilibrium is obtained substituting the fields in eq. \eqref{weakarbi} by finite dimensional fields, defined as the interpolation of the value of the fields at the center of the voxels using trigonometric polynomials \citep{Geers2017}. Expressing the convolution operation in  eq. \eqref{weakarbi}  in the Fourier space and integrating provides a discrete expression for the linear momentum balance 

\begin{equation}
\mathcal{G}^*\left(\mathbf{P}^{(h)}\left(\mathbf{F}^{(h)},\boldsymbol{\alpha}\right)\right)=\mathbf{0}^{(h)} \label{eql_discrete1}
\end{equation}

\noindent where $\mathcal{G}^*$ is a linear map that acts on a discrete voxel tensor field $\mathbf{P}^{(h)}$ defined in $\mathbb{R}^{9N^3}$ to create a new voxel tensor field defined in the same space. The linear map represents a convolution in the real space and is computed as
\begin{equation}
\mathcal{G}^*\left(\mathbf{P}^{(h)}\right):=\mathcal{F}^{-1}\left\{ \widehat{\mathbb{G}^*}:\mathcal{F} \left\{\mathbf{P}^{(h)}\left(\mathbf{F}^{(h)},\boldsymbol{\alpha}\right) \right\} \right\}=\mathbf{0}^{(h)}
\label{eql_discrete2}
\end{equation}

\noindent where $\mathcal{F}$ and $\mathcal{F}^{-1}$ stand for the discrete Fourier transform and its inverse and $\widehat{\mathbb{G}^*}$ is the projection operator in the Fourier space, a fourth order tensor field defined by \cite{Geers2017} for an imposed deformation gradient and extended by \cite{LS2019b} for generalized loading conditions.

The derivatives involved in the definition of the Fourier space operator $\widehat{\mathbb{G}^*}$ as well as in the evaluation of the Nye tensor, eq. \eqref{Nye_finite_strain}, might lead to numerical instabilities (Gibbs oscillations) when applied to fields with non-smooth variations. To overcome this type of numerical problem, \emph{discrete differential operators} were first introduced in \cite{Mueller98}. They replace the partial derivatives of the differential operators in the real space by a finite difference derivation rule. \citep{BERBENNI2014} extended this idea to derive first and second order differential operators based on central differences for Poisson and Navier equations, and this scheme was used in \cite{LN16} to treat the second gradient of plastic distortion. An alternative finite difference scheme was proposed in \cite{W15}, the rotated scheme, which was the one selected in this investigation for both the definition of the projection operator $\widehat{\mathbb{G}^*}$ and the definition of the Nye tensor. Indeed, \cite{DJAKA2019} also used recently this definition of the differential operators to compute the Nye tensor with very good accuracy. The discrete derivative rules used to obtain $\widehat{\mathbb{G}^*}$ and to compute $\nabla \times \mathbf{G}$ are detailed in  \ref{App0} while the numerical details of the  evaluation of $\mathbf{G}$ using FFT and a benchmark example to show the accuracy of the procedure are reported in  \ref{App1}.

 Eq. \eqref{eql_discrete2} is an algebraic system of $9N^3$ non-linear equations in which the unknown is the value of the deformation gradient at each voxel $\mathbf{F}^{(h)}$. The macroscopic load enters in the system by splitting the components of the tensors in eq. \eqref{eql_discrete2}  in which the macroscopic data is given into its average part (given) and fluctuation (unknown) and using the corresponding definition projector operator, $\widehat{\mathbb{G}^*}$ for those components \citep{LS2019b}.

The non-linear system defined in eq. \eqref{eql_discrete2} is solved by splitting the load history in $n$ increments. The unknown at the end of each increment $k$ is $\mathbf{F}^{(h)}_k$, the deformation gradient field corresponding to time $t_k$.  For simplicity, the discrete fields $\mathbf{F}^{(h)}_k$ and $\mathbf{P}^{(h)}_k$ will be named $\mathbf{F}_k$ and $\mathbf{P}_k$. The velocity gradient during increment $k$ is obtained as
$\mathbf{L}_k=\frac{1}{t_{k}-t_{k-1}}(\mathbf{F}_k-\mathbf{F}_{k-1})\mathbf{F}_k^{-1}$.
The non-linear system at time $t_k$ is then solved using using the Newton-Raphson method. The linear system,  whose solution provides the correction of the displacement, $\delta \mathbf{F}$, at each iteration $i$,  is given by

\begin{equation}\label{linequil2ast}
\mathcal{G}^\ast\left(\mathbb{K}:\delta \mathbf{F}\right)=-\mathcal{G}^\ast\left(\mathbf{P}\left(\mathbf{F}^{i}\right)-\overline{\mathbf{P}}_k\right)
\end{equation}

\noindent where $\overline{\mathbf{P}}_k$ is a tensor containing the non-zero $IJ$ components of the imposed stress at time $t_k$ and $\mathbf{F}^{i}=\overline{\mathbf{F}}_{k}+\delta \mathbf{F}^1+\delta \mathbf{F}^2\cdots+\delta \mathbf{F}^{i-1}$. The linear system, eq. \eqref{linequil2ast}, is solved using the conjugate gradient method. The Newton-Raphson iterations for increment $k$ are finished when the correction $\delta \mathbf{F}$ is below a given tolerance.

\section{Effect of the grain size on the flow stress}\label{sec3}

The effect of the grain size on the  tensile behavior of Cu, Al, Ag and Ni polycrystals was analyzed using the computational homogenization strategy and the SGCP model presented above. Four different simulations were carried out for each material, with average grain sizes $\overline{D}_g$ of 10, 20, 40, and 80 $\mu$m.  One additional simulation was carried out for each material in which the effect of the GNDs on the mean free path and the flow stress was not included. Thus, these simulations do not include any length scale and provide the effective behavior of the polycrystal for an "infinite" average grain size. All the simulations were performed at a constant strain rate of 10$^{-4}$ s$^{-1}$ under uniaxial tension up to an applied strain of 5\%. 

The physical parameters of the SGCP were identified from either experiments or discrete dislocation dynamics simulations at lower length scales in the literature. There are a number of parameters common to all FCC metals (strain rate sensitivity as well as the interaction coefficients, $q^{\alpha\beta}$, among dislocations) that are depicted in Table \ref{table:1}. The first set of physical parameters which depended of the particular FCC material were the elastic constants and the Burgers vector,  that can be found in the literature \citep{Al-elastic-constants, Ni-elastic-constants, Ag-elastic-constants}. They are given in Table \ref{table:2}, together with the shear modulus $\mu$ parallel to the slip plane, which can be computed from the elastic constants. The  similitude coefficient $K$ is a dimensionless constant that arises from the  experimental observation (known as the  similitude principle) that relates the flow stress $\tau$ with the average wavelength of the characteristic dislocation pattern $d$ according to $\tau = K \mu b/d$ in the case of FCC metals. The actual values of the similitude coefficient $K$ were identified by \cite{HSL18} and \cite{RHL19} for Cu , Al, Ni and Ag from the experimental data in \cite{SK11}, and are given in Table \ref{table:2}. Finally, the effective annihilation distance, $y_c$, was assumed to be the average between the mean annihilation distance of edge dislocations ($\approx 6b$ \citep{MWW00}) and screw dislocations. The latter depends on the ability of  screw dislocations to cross-slip, a thermally-activated phenomenon which depends on the stacking fault energy of the crystal and also on temperature, strain rate and applied stress. The annihilation distance for screw dislocations under quasi-static loading conditions and ambient temperature was estimated by \cite{RHL19} for the FCC metals considered in this investigation, and the resulting effective annihilation distance can also be found in Table \ref{table:2}.

\begin{table}[h!]
\caption{Parameters of the SGCP model common to all FCC crystals.}
\centering
\begin{tabular}{| l| c|}
 \hline
 {\it Visco-plastic parameters}  \citep{vicoplastic-parameters-fcc} & \\
 Reference shear strain rate $\dot{\gamma}_0$  (s$^{-1}$) & 0.001\\
 Strain rate sensitivity coefficient $m$ & 0.05\\
 \hline
 {\it Interaction coefficients:} \citep{interaction-matrix-coefficients} & \\
 Self interaction  & 0.122\\
 Coplanar interaction & 0.122\\
 Collinear interaction & 0.657\\
 Hirth lock & 0.084\\
 Glissile junction & 0.137\\
 Lomer-Cottrell lock & 0.118\\
 \hline
\end{tabular}
\label{table:1}
\end{table}

\begin{table}[h!]
\caption{Parameters of the SGCP model for Cu, Al, Ni and Ag.}
\centering
\begin{tabular}{| l| c|c| c| c | }
 
 \hline
 & Cu & Al & Ni & Ag\\
  \hline
 {\it Elastic constants} (GPa) &&&&\\
 $C_{11} $ & 169& 108   & 249 & 124\\
 $C_{12}$ & 122& 61.3  & 155  & 93.7\\
 $C_{44} $ & 75.3 & 28.0 & 114  & 46.1 \\
 $\mu $  & 30.5 & 25.0  & 58.4   & 19.5 \\
 \hline
 \it{Dislocation parameters} & & & & \\
 Burgers vector $b$ (nm) & 0.256& 0.286  & 0.250  & 0.288  \\
Effective annihilation distance $y_c$  (nm) &15& 56  & 14& 12.5  \\
 Similitude coefficient K & 6& 9  & 11& 5\\
 \hline
\end{tabular}
\label{table:2}
\end{table}

\begin{figure}[h!]
    \centering
    \includegraphics[scale= 0.8]{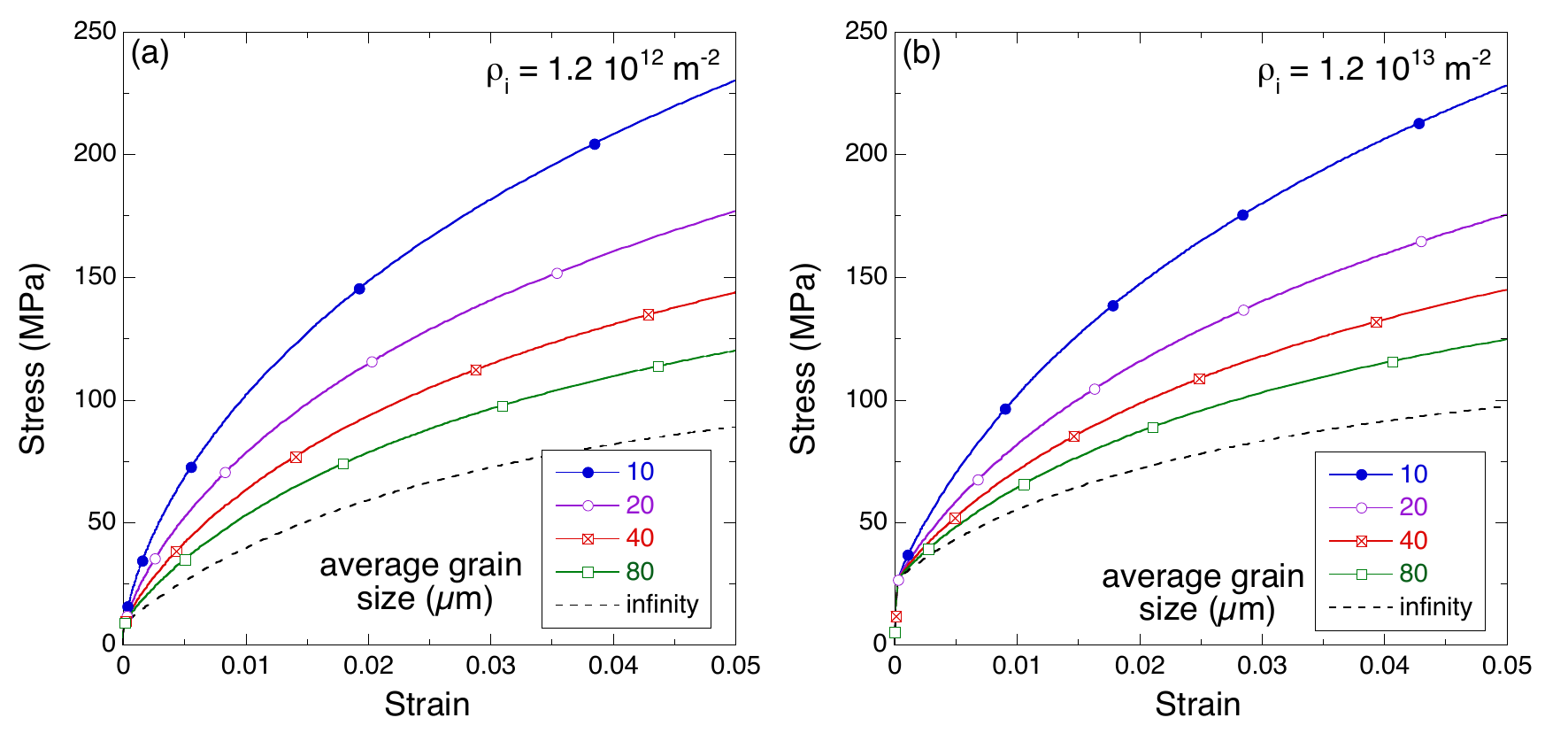}
     \includegraphics[scale= 0.8]{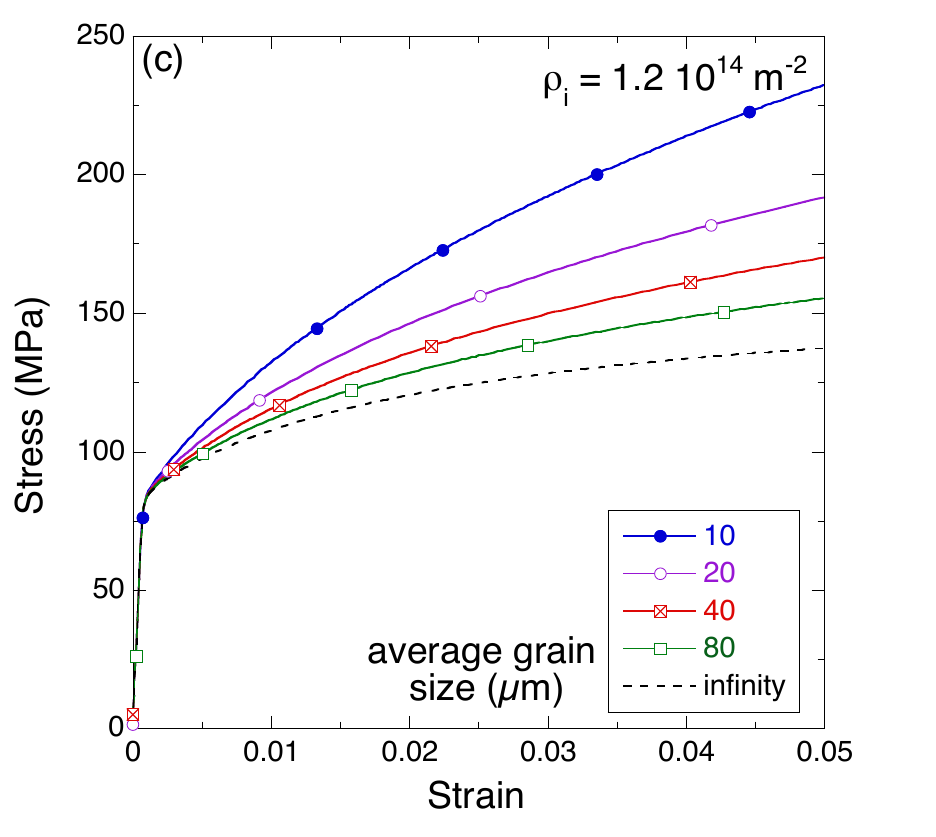}
    \caption{Stress-strain curves of Cu polycrystals as a function of the average grain size. 
(a) Initial dislocation density, $\rho_i$ =1.2 10$^{12}$ m$^{-2}$. (b) Initial dislocation density, $\rho_i$ =1.2 10$^{13}$ m$^{-2}$. (c) Initial dislocation density, $\rho_i$ =1.2 10$^{14}$ m$^{-2}$. The broken lines stand for the results obtained when the contribution of the GNDs is not included in the model.}
    \label{Cu}
\end{figure}

The tensile behavior of Cu polycrystals was computed for three different values of the initial SSD density,  $\rho_i$ =1.2 10$^{12}$ m$^{-2}$, 1.2 10$^{13}$ m$^{-2}$ and 1.2 10$^{14}$ m$^{-2}$. The initial dislocation density in each slip system was 1/12 of the total. The corresponding stress-strain curves are plotted in Figs. \ref{Cu}a, b and c. The broken lines stand for the results obtained when the contribution of the GNDs is not included in the model. The initial yield stress of all the polycrystals is independent of grain size and only depends on the initial SSD density. Afterwards, hardening is observed in all cases and the strain hardening rate increases as the average crystal size decreases. It is worth noting that -although the initial flow stress is highly dependent on the initial SSD density- the flow stress at 5\% only depends weakly on the initial SSD density and is mainly controlled by the average grain size. This behavior indicates that the generation of new SSDs and GNDs during deformation rapidly smooths out the initial differences in SSD densities. 

\begin{figure}[h!]
    \centering
    \includegraphics [width=\textwidth]{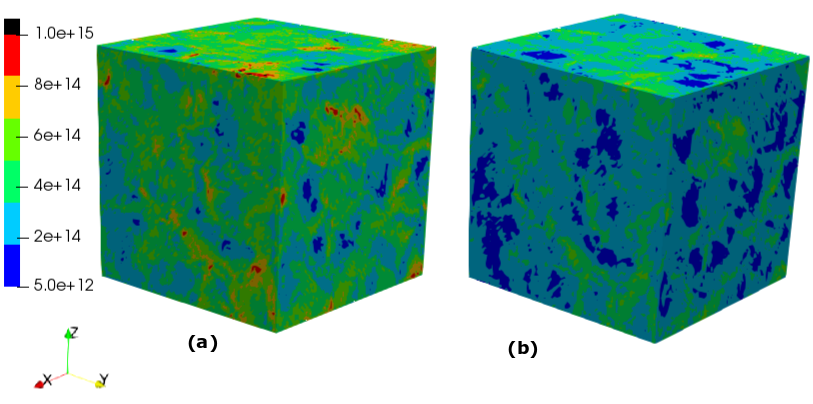}
    \caption{Contour plot of the SSD density in all the slip systems of Cu polycrystals. (a) $\overline{D}_g$ = 10$\mu$m. (b) $\overline{D}_g$ = 40 $\mu$m.  The far-field applied strain was 5\% and the initial dislocation density $\rho_i$ =1.2 10$^{12}$ m$^{-2}$. Dislocation densities are expressed in m$^{-2}$.}
    \label{SSD}
\end{figure}

\begin{figure}[h!]
    \centering
    \includegraphics[width=\textwidth]{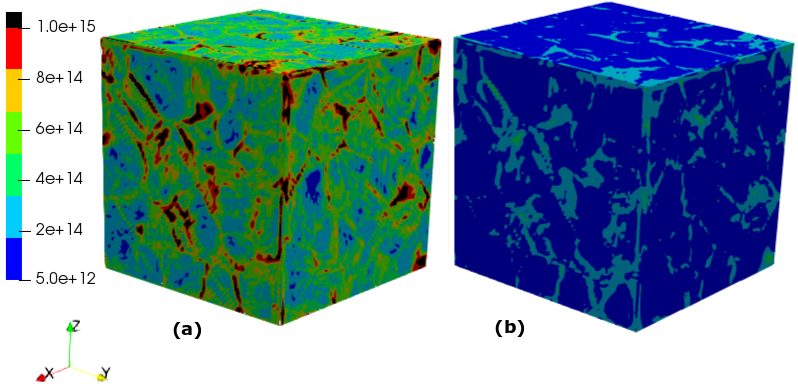}
    \caption{Contour plot of the GND density in all the slip systems of Cu polycrystals. (a) $\overline{D}_g$ = 10$\mu$ m. (b) $\overline{D}_g$ = 40 $\mu$ m  The far-field applied strain was 5\% and the initial dislocation density $\rho_i$ =1.2 10$^{12}$ m$^{-2}$. Dislocation densities are expressed in m$^{-2}$.}
    \label{GND}
\end{figure}

The dislocation densities of SSDs and GNDs of Cu polycrystals deformed up to 5\% are shown in the contour plots of Figs. \ref{SSD} and \ref{GND}, respectively. Each figure includes the contour plots of polycrystals with $\overline{D}_g$ = 10 and 40 $\mu$m with an initial dislocation density $\rho_i$ =1.2 10$^{12}$ m$^{-2}$. In the case of the polycrystals with $\overline{D}_g$ = 40 $\mu$m, the SSD density (Fig. \ref{SSD}(b)) is higher than the GND density (Fig. \ref{GND}(b)), indicating that most of the latent hardening for this grain size was provided by the interaction of the SSDs in the bulk of each grain. On the contrary, the GND density (Fig. \ref{GND}(a)) was much higher than the SSD density (Fig. \ref{SSD}(a)) in the polycrystal with $\overline{D}_g$ = 10 $\mu$m and the GNDs were partially responsible for the large strain hardening. It should also be noted that the density of the SSDs was also higher in the polycrystal with $\overline{D}_g$ = 10 $\mu$m than in the one with $\overline{D}_g$ = 40 $\mu$m because the GNDs also contributed to reduce the mean-free path (see eq. \eqref{eq:meanfree}) and, thus, to increase the rate of generation of new SSDs during deformation. Finally, the GNDs were mainly accumulated at the grain boundaries (Fig. \ref{GND}) but not all the boundaries presented a large density of GNDs because it depended on the incompatibility of deformation between the grains at each boundary.

\begin{figure}[h!]
    \centering
    \includegraphics[width=\textwidth]{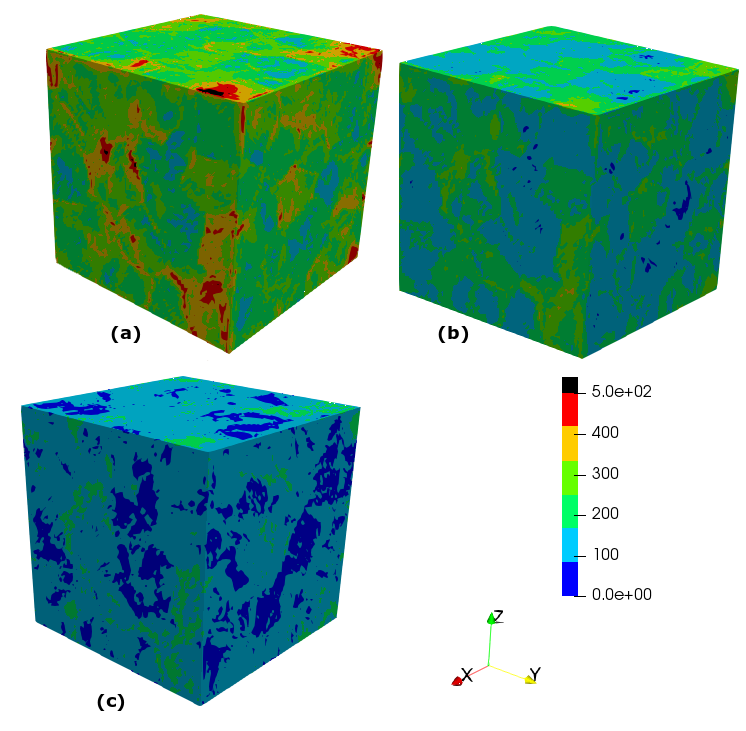}
    \caption{Contour plots of the Von Misses stress in Cu polycrystals. (a) $\overline{D}_g$ = 10$\mu$ m. (b) $\overline{D}_g$ = 40 $\mu$m. (c)  $\overline{D}_g$ = infinity. The far-field applied strain was 5\% and the initial dislocation density $\rho_i$ =1.2 10$^{12}$ m$^{-2}$. Stresses are expressed in MPa}
    \label{VM}
\end{figure}

The effect of the different dislocation densities on the Von Mises stress is shown in the contour plots in Fig. \ref{VM} for the polycrystals with $\overline{D}_g$ = 10 $\mu$m, 20 $\mu$m and infinity. They were fairly homogeneous and low in the polycrystal with "infinite" grain size (Fig. \ref{VM}(c)) and the heterogeneity in the stress distribution as well as the average stress level increased as the grain size decreased from 40 $\mu$m (Fig. \ref{VM}(b)) to 10 $\mu$m (Fig. \ref{VM}(a)). 

\begin{figure}[h!]
    \centering
    \includegraphics [scale=0.8]{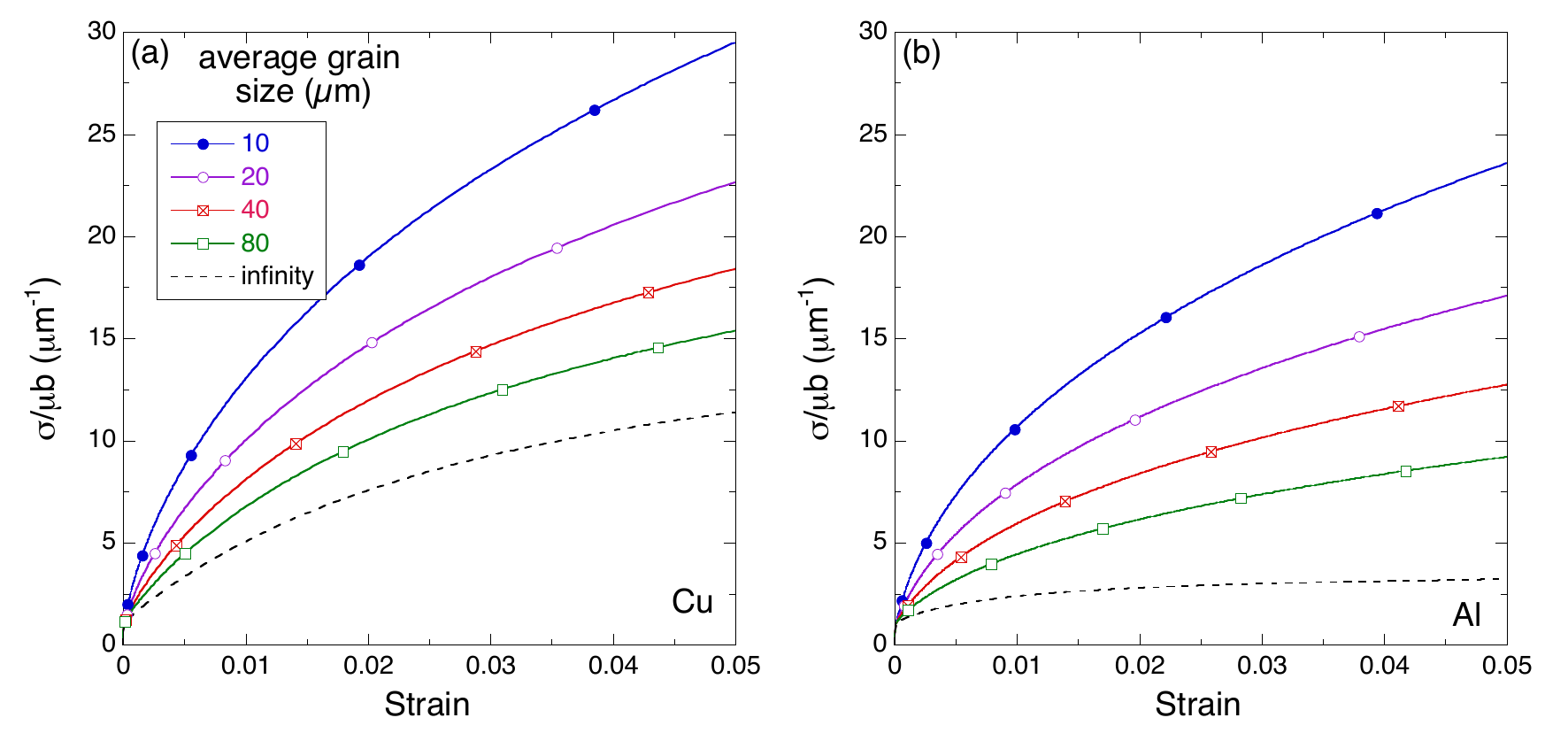}
       \includegraphics [scale=0.8]{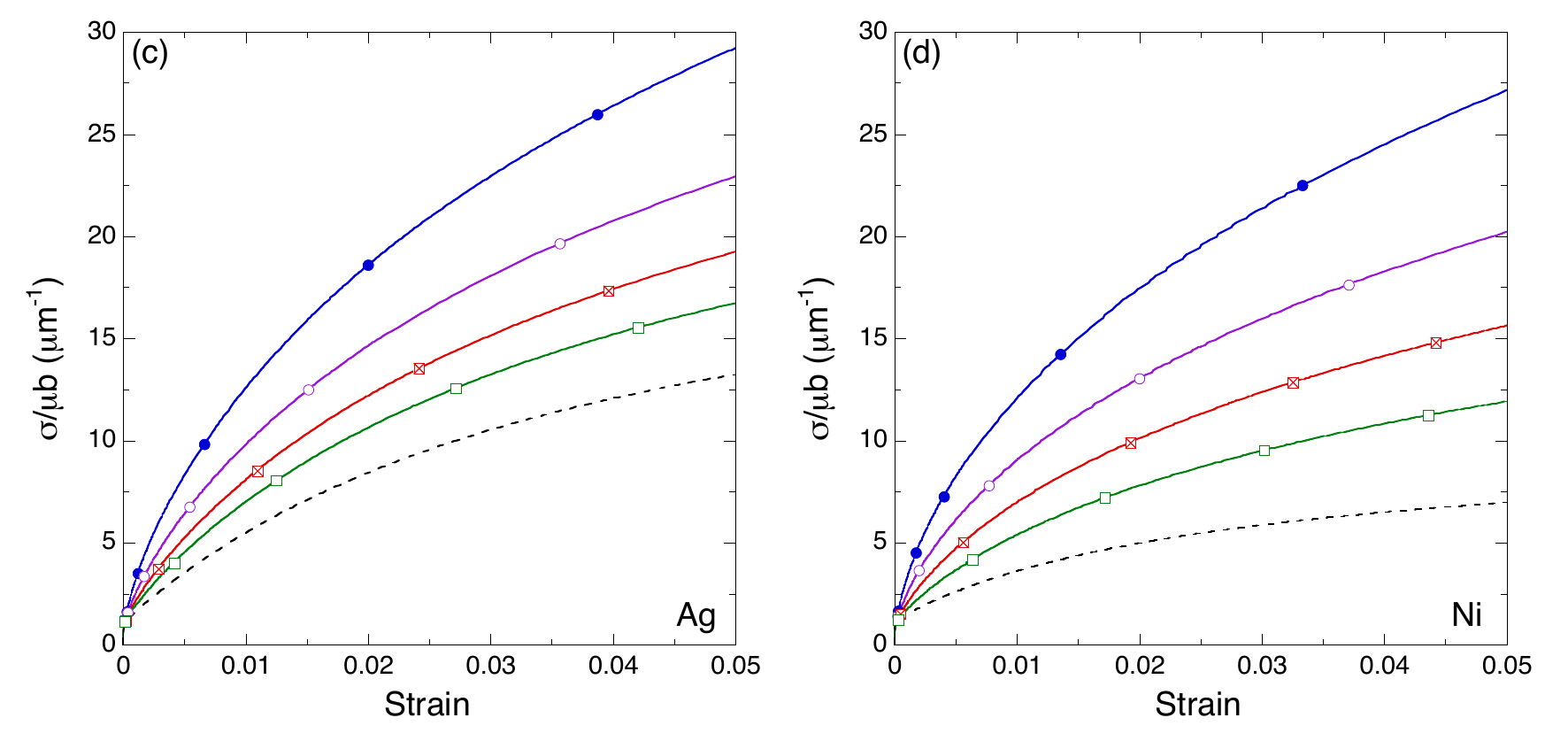}
    \caption{Stress-strain curves of different FCC polycrystals as a function of the average grain size. (a) Cu, (b) Al, (c) Ag and (d) Ni. The initial dislocation density was $\rho_i=1.2 \, 10^{12} m^{-2}$ in all cases. The broken lines stand for the results obtained when the contribution of the GNDs is not included in the model. The stresses were normalized by $\mu b$ in all cases.}
    \label{FCC}
\end{figure}

In order to understand the effect of grain boundaries on the flow stress of different FCC metals, simulations were also carried in Al, Ag and Ni polycrystals with the same initial dislocation density on each slip system, 10$^{11}$ m$^{-2}$, leading to a total initial dislocation density of $\rho_i$ =1.2 10$^{12}$ m$^{-2}$. The corresponding stress-strain curves are plotted in Fig. \ref{FCC}a to d for Cu, Al, Ni and Ag. The stress were normalized by $\mu b$ in all cases to show more clearly the differences among FCC polycrystals as a result of interaction, accumulation, and annihilation of dislocations. 

The broken lines in Fig. \ref{FCC} stand for the results obtained when the contribution of the GNDs is not included in the model and they show the interplay between the generation of dislocations in the bulk (controlled by the dislocation mean free path and the similitude coefficient $K$) and the annihilation of dislocations (controlled by the effective annihilation distance $y_c$) for different FCC metals. Ag and Cu (that have the smallest similitude coefficients and effective annihilation distances, Table \ref{table:2}) show the largest strain hardening due to the rapid accumulation of dislocations in the bulk (Figs. \ref{FCC}a and c). On the contrary, the strain hardening rate of Al (Fig. \ref{FCC}b) becomes close to 0 at very low strains ($<$ 2\%). Al presents the largest critical annihilation distance  (due to the large stacking fault energy, that favours cross slip) and the similitude coefficient is also high (Table \ref{table:2}). Thus, the generation of new dislocations in the bulk is rapidly balanced with the annihilation of dislocations reaching a steady-state regime.

The deformation incompatibility between grains leads to the accumulation of GNDs at the grain boundaries and increases the strain hardening rate in all cases. Moreover, the GND densities increase as the average grain size decreases, leading to the development of a Hall-Petch behavior. GNDs contribute to the strain hardening in two ways. Firstly, they reduce the mean free path near the grain boundaries (eq. \eqref{eq:meanfree}) and, thus, increase the generation rate of new SSDs. Secondly, they contribute directly to the hardening through the Taylor equation, eq. \eqref{eq:Taylor_hard}. It is important to notice that the GND density (for a given polycrystal) increases monotonically with the applied strain to maintain the kinematic compatibility between grains. Thus, the SGCP model induces a continuous hardening of the polycrystal and the steady-state situation in which the dislocation accumulation rate is zero is never attained. This behavior is contrary to the experimental evidence which shows that dynamic recovery due to dislocation annihilation becomes dominant when the applied strain is large enough \citep{KM81}.

Of the two hardening mechanisms indicated above, the latter is proportional to $\mu b$ (eq.Ê\eqref{eq:Taylor_hard}) and should not lead to any difference between the FCC metals in Fig. \ref{FCC}, in which the stresses are normalized by $\mu b$. The reduction in the mean free path due to the development of GNDs, eq. \eqref{eq:meanfree}, depends on the similitude coefficient $K$ and, thus, the strain hardening rate of Cu and Ag polycrystals is higher than that of Al and Ni for the same average grain size.

\subsection{Scaling laws for the flow stress}

Following the pioneer work of Hall and Petch, many authors have reported  that most of the data on the experimental effect of the grain size of the flow stress of metallic polycrystals could be better approximated by a generalized form of the Hall-Petch law \citep{RP86},

\begin{equation}
(\sigma_y-\sigma_\infty) = C \overline{D}_g^{-x}
\label{eq:x}
\end{equation}

\noindent where $\sigma_y$ is the flow stress, $\sigma_\infty$ the flow stress for a polycrystal with "infinite" grain size,  and $C$ and $x$ two material constants, the latter in the range $ 0 < x \le 1$ \citep{RP86}. However, it has been recently shown that this law breaks down for FCC polycrystals with large initial dislocation densities ($>$ 10$^{14}$ m$^{-2}$) and grain sizes ($>$ 40 $\mu$m) \citep{HSL18}. This result obtained using crystal plasticity is in agreement with theoretical results \cite{ZS14} and dislocation dynamics simulations \citep{E15} as well as with recent analysis of the experimental data \citep{DR19}, which indicate that the strengthening associated with the grain size in plastic polycrystals, $\sigma_y-\sigma_\infty$, has to be expressed as

\begin{equation}
\sigma_y - \sigma_\infty = \sigma_\infty \Delta (\overline{D}_g\sqrt{\rho})
\label{eq:SizeEffect}
\end{equation}

\noindent where $\Delta (\overline{D}_g\sqrt{\rho})$ is a non-dimensional function of the ratio between two length scales: the average grain size ($\overline{D}_g$) and the average dislocation spacing ($1/\sqrt{\rho}$).

\begin{figure*}[h!]
    \centering
        \includegraphics[scale=0.8]{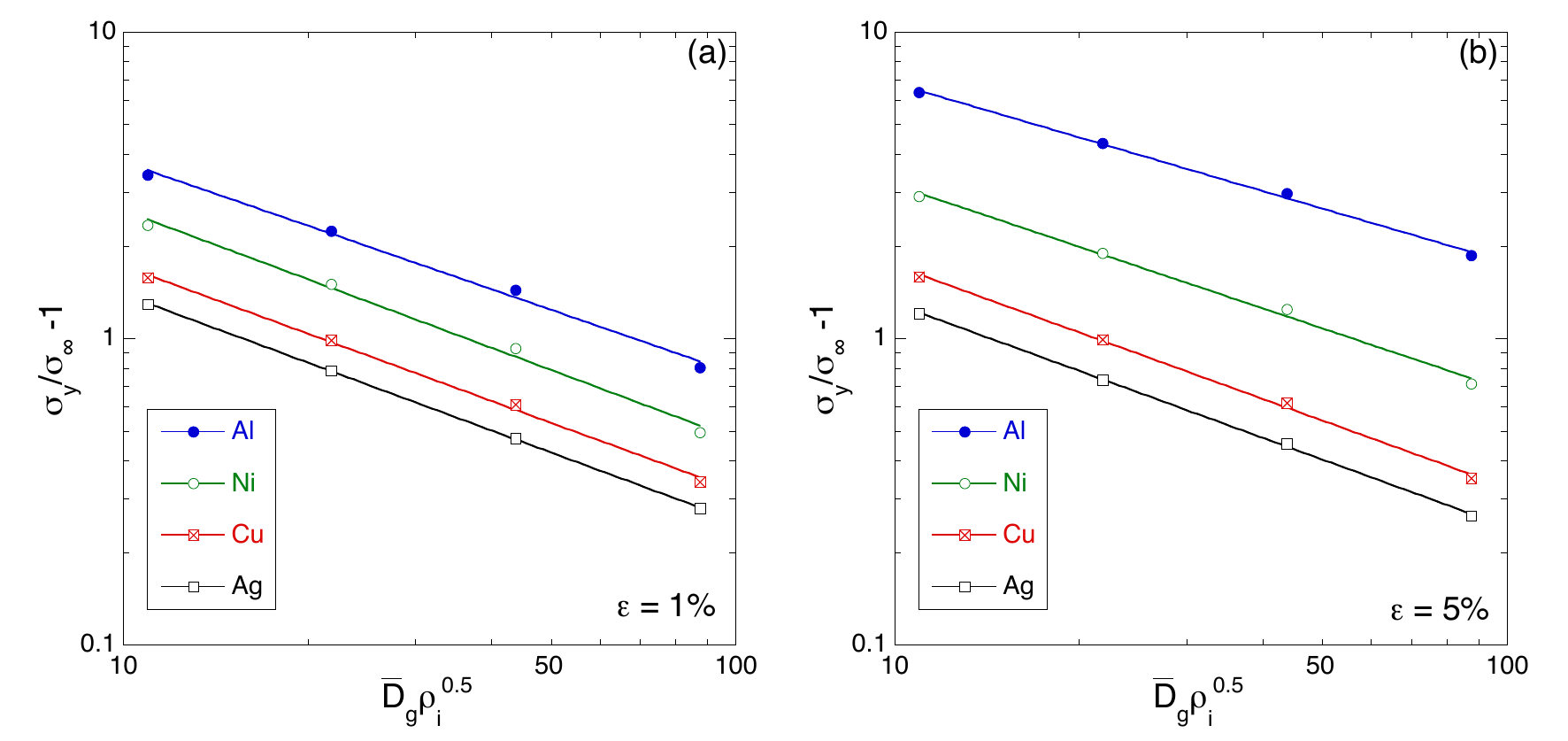}
        \caption{Grain boundary strengthening in FCC metals (Ni, Ag, Al and Cu) as a function of the dimensionless parameter $\overline{D}_g \sqrt{\rho_i}$ . (a) $\epsilon$ = 1\%. (b) $\epsilon$ = 5\%. The points represent the result of the simulations and the lines correspond to the fitting to the linear relation defined in eq. (\ref{eq:SizeEffect2}) } \label{figFCC}
\end{figure*}

The correlation of the grain boundary strengthening obtained in the simulation of the different FCC polycrystals respect the eq. \eqref{eq:SizeEffect} has been checked in Fig. \ref{figFCC}. In this figure, the strengthening induced by the grain boundaries in Cu, Al, Ni and Ag, defined as $\sigma_y/\sigma_\infty -1$, has been plotted as a function of the dimensionless parameter $\overline{D}_g\sqrt{\rho_i}$ for applied strains $\epsilon$  = 1\% and 5\%. The initial dislocation density in all cases was $\rho_i$ =1.2 10$^{12}$ m$^{-2}$. The results of the SGCP simulations of polycrystals can be accurately fitted to an expression of the form

\begin{equation}
\sigma_y/\sigma_\infty -1 =  C (\overline{D_g}\sqrt{\rho_i})^{-x}
\label{eq:SizeEffect2}
\end{equation}

\noindent where the exponent $x$ varied from 0.69 in Al to 0.74 in Ag for $\epsilon$ = 1\% and from 0.59 in Al to 0.73 in Ag at $\epsilon$ = 5\%. Thus, $x$ was similar for all FCC polycrystals and  weakly dependent on the applied strain. These similarities in the exponent $x$ could be expected because the hardening mechanism due to the presence of grain boundaries in FCC polycrystals is the same (the generation of GNDs) and it is controlled by the kinematics of deformation.

The different values of $C$ in eq. \eqref{eq:SizeEffect2} illustrate the different strengthening capability of grain boundaries for FCC polycrystals. The density of GNDs for a given average grain size is determined by the applied strain and it is independent of the type of material (if elastic anisotropy is neglected in comparison with plastic anisotropy). Thus,  the main contribution of the GNDs to the strength is given by the Taylor equation, eq. \eqref{eq:Taylor_hard}, while the contribution associated with the reduction in the mean free path, eq. \eqref{eq:meanfree}, which depends on $K$ is smaller. As a result, grain boundary strengthening is more important in FCC polycrystals which show limited strain hardening in the absence of grain boundaries (Al) and minimum in the case of Cu and Ag, which show  marked strain hardening even in polycrystals with very large grain size.

\section{Comparison with experiments}

In order to assess the validity of the simulations presented above, the results of the numerical simulations were compared with experimental data in the literature for Cu \citep{ACD62, HR82}, Al \citep{al-exp-dat}, Ag \citep{ag-exp-dat} and Ni \citep{ni-exp-dat}. The results of the polycrystal homogenization simulations were obtained using the parameters in Tab. \ref{table:1} and Tab. \ref{table:2} and an initial dislocation density of 1.2  10$^{12}$ m$^{-2}$, which corresponds to a well-annealed polycrystal. All the simulations were performed under quasi-static strain rates ($\approx 10^{-4}\ s^{-1}$).

\begin{figure}[h!]
    \centering
    \includegraphics[scale=0.8]{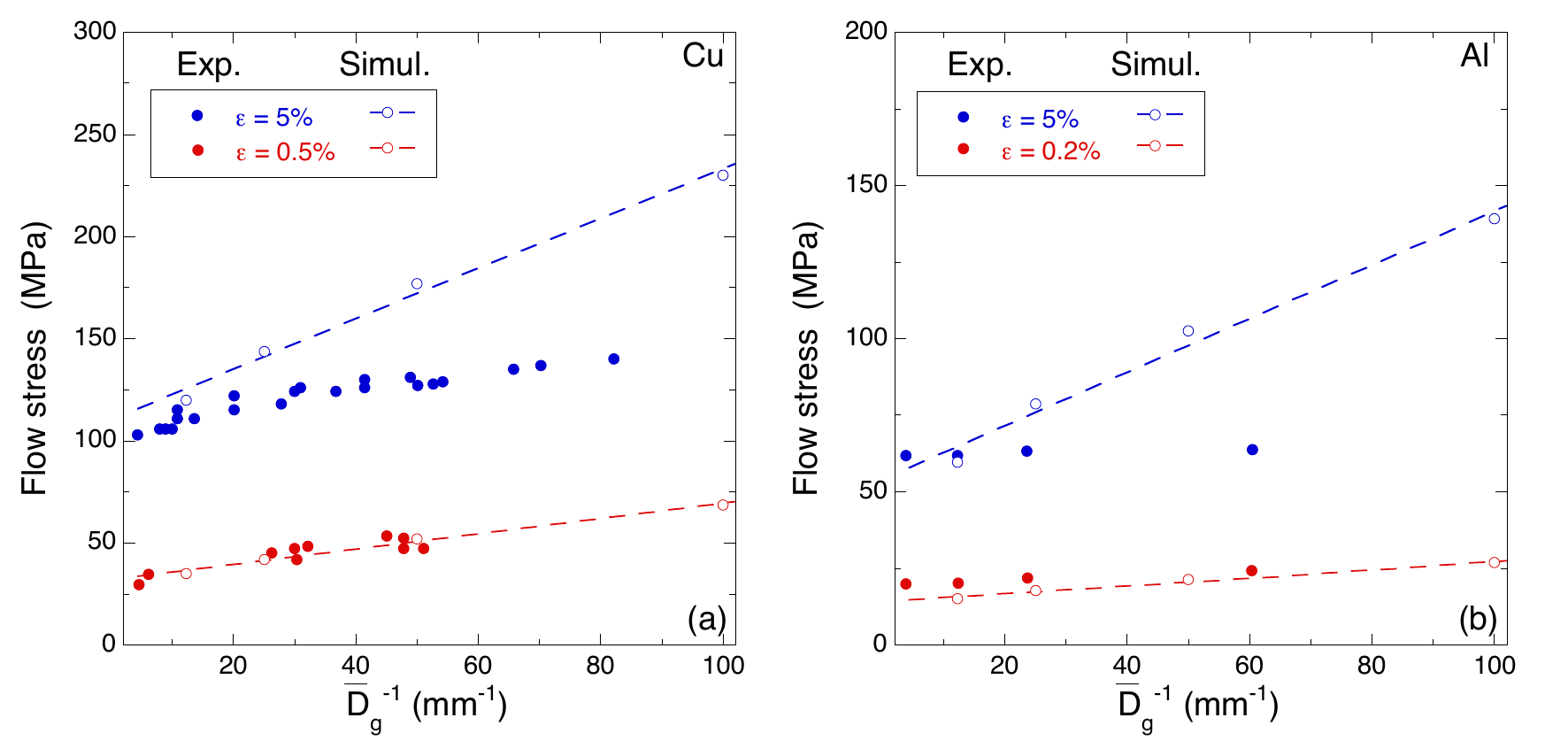}
     \includegraphics[scale=0.8]{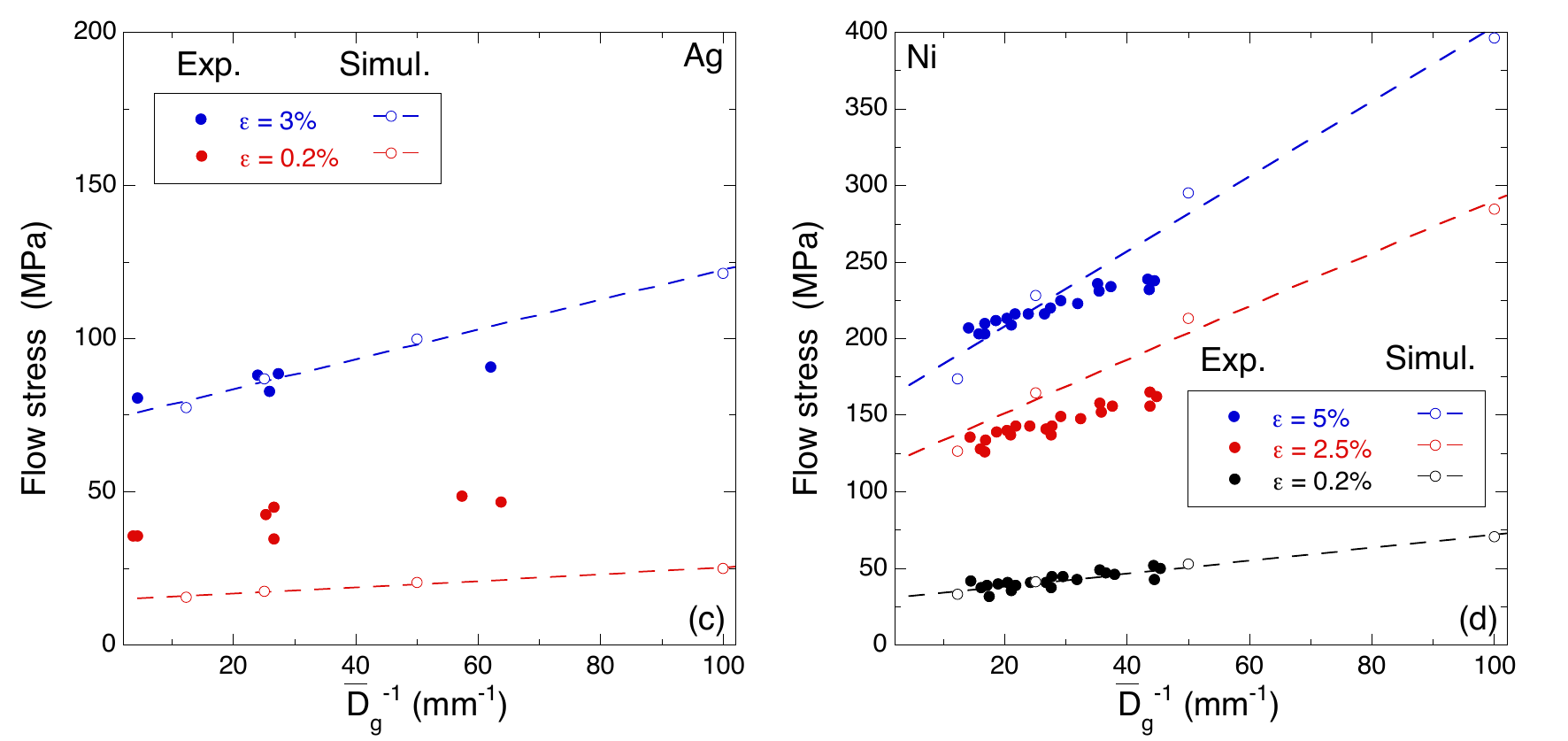}
    \caption{Influence of grain size on the flow stress of polycrystalline FCC metals. Experimental results for different values of the applied strain and predictions of the SGCP model. (a) Cu. (b) Al, (c) Ag. (d) Ni. See text for more details.}
    \label{HPexpsim}
\end{figure}

The experimental values of the flow stress and the corresponding results obtained from the numerical simulations for different values of the applied strain $\epsilon$ are plotted in Figs. \ref{HPexpsim}a, b, c, and d for Cu, Al, Ag and Ni polycrystals, respectively. The flow stress is plotted {\it vs.} $\overline{D}_g^{-1}$ because recent analyses indicate that the experimental data of flow stress follows $\overline{D}_g^{-1}$ rather than  $\overline{D}_g^{-0.5}$ \citep{DB13, DB14} but the conclusions below are independent on whether the flow stress is plotted {\it vs.}  $\overline{D}_g^{-1}$ or  $\overline{D}_g^{-0.5}$.

The comparison between experiments and simulations in Fig. \ref{HPexpsim} shows a good agreement between both for small values of the applied strain ($\epsilon \le $ 0.5\%) for all polycrystals with exception of Ag. On the contrary, the flow stress predicted by the simulations tends to overestimate the experimental flow stress for large strains ($\epsilon$ = 5\%) and small grain sizes ($\overline{D}_g \le$ 20 $\mu$m or $\overline{D}_g^{-1} \ge$ 50 mm$^{-1}$). The differences are obvious in the case of Cu, Al and Ag and there are not experimental data available in this range in the case of Ni. Nevertheless, if the available experimental results in Ni are extrapolated to the small grain size regime, it also seems that the simulation results overestimate the flow stress. It should be noted, however, that the computational homogenization strategy in combination with the SGCP model provides a remarkably good approximation to the experimental data without any adjustable parameters. In particular, the model parameters that control the interaction among dislocations and the generation and annihilation of dislocations (Tables \ref{table:1} and \ref{table:2}) come from dislocation dynamics simulations or independent experimental observations and the  only intrinsic length scale in the model is the Burgers vector of each FCC metal.

The SGCP model assumes that the hardening due to the grain boundaries is negligible at very low strains because there are not GNDs at the onset plastic deformation. Thus, the flow stress at small strains is fairly independent of the grain size and controlled by the initial density of SSDs, which is not known for the experimental data. This may explain the differences between the experimental and simulated flow stress in Ag at $\epsilon$ = 0.2\% that may due to a gap between the experimental and simulated initial densities of SSDs (Fig. \ref{HPexpsim}c). In fact, the differences in the flow stress between experiments and simulations disappeared at $\epsilon$ = 3\% because of the rapid generation of SSDs and GNDs with strain.

The differences between experiments and simulations for small grain sizes and large strains are the result of the hardening hypothesis in the SGCP model. According to this model, the density of GNDs is controlled by the lattice curvature around the grain boundaries due to the inhomogeneity of the plastic deformation and increases continuously with the applied strain. As the flow stress is proportional to the square root of the dislocation density, the model predicts continuous strain hardening, leading to the very high values of the flow stress for large strains and small grain sizes, where the contribution of GNDs is dominant over the one from SSDs. It should be noted that this large strain hardening at large strains is not found in the mechanical tests of polycrystals \citep{HR82, ni-exp-dat, ag-exp-dat, al-exp-dat} where dynamic recovery leads to very low strain hardening rates (similar to the ones found in Fig. \ref{FCC} for polycrystals with "infinite" grain size) for large strains. 

In another investigation \citep{RHL19}, the effect of grain size on the flow strength of FCC polycrystals has been analyzed using computational homogenization in combination with a standard, dislocation-based crystal plasticity model which only considers SSDs. In this approach, the influence of grain boundaries is taken into account by modifying the Kocks-Mecking equation, eq. \eqref{eq:kocks}, so the rate of generation of new dislocations near the grain boundaries is inversely proportional to the distance to the grain boundary. This lead to an increase in the dislocation density near the grain boundaries, mimicking the formation of dislocation pile-ups near the boundaries. Nevertheless, dislocation annihilation is included in the model and the maximum dislocation density near the grain boundaries is limited. This is not the case for the SGCP model, where the density of GNDs is not affected by the annihilation distance. As a result, the predictions of the simulations in \citep{RHL19} are in better agreement with the experimental data for large strains and small grain sizes although they also tend to overestimate the flow stress under these conditions.

Another reason why the flow strength is overestimated by the SGCP model can be found in the presence of slip transfer between neighbour grains. It has been reported experimentally that dislocations moving in one slip system in one grain can easily move into another slip system in the adjacent grain if the Luster-Morris parameter $m' = cos \psi \cos \kappa \ge 0.95$ in FCC Al, where $\psi$ and $\kappa$ are the angles between the slip plane normals and the Burgers vector directions, respectively, for each pair of slip systems \citep{BAP19}. Slip transfer helps to accommodate the heterogeneous deformation between neighbour grains without generating any GNDs. Although slip transfer seems to be limited to neighbour grains with small misorientation  ($\approx < 20^\circ$) \citep{BAP19}, this mechanism will reduce the actual strengthening induced by the grain boundaries, particularly in polycrystals with small grain size and deformed up to large strains.

\section{Conclusions}

The influence of grain size on the flow stress of various FCC polycrystals (Cu, Al, Ag and Ni) has been ascertained by means of computational homogenization using the FFT of a representative volume element of the microstructure in combination with a strain gradient crystal plasticity model. The density of geometrically necessary dislocations resulting from the incompatibility of deformation among different crystals was obtained from the Nye tensor, which was efficiently obtained from the curl operation in the Fourier space. All the parameters of the crystal plasticity model have a clear physical meaning and were obtained for each FCC polycrystal from either dislocation dynamics simulations or experimental data in the literature and the only length scale in the simulations was the Burgers vector.

The simulation results were able to reproduce the effect of grain size on the flow stress due to the generation of geometrically-necessary dislocations around the grain boundaries. Moreover, they corroborated previous results that indicate that the overall strengthening due to grain boundaries is a function of the dimensionless parameter $\overline{D}_g \sqrt{\rho_i}$ where $\overline{D}_g$ is the average grain size and $\rho_i$ the initial dislocation density. The simulation predictions of the flow stress were in good agreement with the experimental data for Cu, Al, Ag and Ni polycrystals for grain sizes $>$ 20 $\mu$m and strains $ <$ 5\% and provided a physical explanation for the higher strengthening provided grain boundaries in Al and Ni, as compared with Cu and Ag. Nevertheless, the simulations tended to overestimated the flow strength for smaller grain sizes and larger strains because the model predicts a continuous hardening as a result of continuous accumulation of geometrically-necessary dislocations at the grain boundaries, which is not compensated by annihilation of dislocations due to dynamic recovery. Moreover, slip transfer at low angle grain boundaries is not considered in strain gradient crystal plasticity framework.

In summary, the whole approach shows how the combination of FFT with strain gradient crystal plasticity can be used to include effect of grain boundaries in the mechanical behavior of polycrystals using realistic representative volume elements of the microstructure. This information is important to develop new models of the mechanical behavior of polycrystals that account for the nucleation of damage at the grain boundaries (fatigue, creep, etc.)

\section{Acknowledgments}
This investigation was supported by the European Research Council under the European Union's Horizon 2020 research and innovation programme (Advanced Grant VIRMETAL, grant agreement No. 669141) and the Spanish Ministry of Science through the project ENVIDIA, RTC-2017-6150-4.

\appendix
\section{Discrete derivative rules}\label{App0}
FFT.based approaches to solve partial differential equations are based in the evaluation of the differential operators in the real space using a Fourier transform and a product in the Fourier space. In one dimension
\begin{equation}\label{eq:derivative}
 \widehat{f'(x)}=\mathrm{i} 2\pi (N/L)\xi \hat{f}(\xi).
 \end{equation}

When a periodic domain of size $L$ is introduced and discretized in N equispaced voxels, the values of $\xi$ (for an odd number of voxels) are given by
\begin{equation}\label{freqs}
\xi= \frac{n-(N+1)/2}{N}\quad \text{ for } \quad n=1, \dots, N.
\end{equation}

In this study, the continuum derivate, eq. \eqref{eq:derivative}, is replaced by a finite difference scheme to avoid numerical noise, following  previous investigations \cite{Mueller98,BERBENNI2014,W15}. In particular,  the rotated scheme of \citep{W15} in a 3D domain is used and the directional derivative along direction $i$ is defined using the Fourier space as 
\begin{equation}\label{eq:derivative2}
\widehat{\frac{\partial f}{\partial x_i}}=\mathrm{i} k_i(\boldsymbol{\xi})\ \hat{f}(\boldsymbol{\xi}).
 \end{equation}

\noindent where $k_i(\boldsymbol{\xi})$ are the modified definitions of the frequency vector, obtained following a difference scheme in the real space. The value of $k_i(\boldsymbol{\xi})$ is given by
\begin{equation}\label{kas}
k_i(\xi_1,\xi_2,\xi_3) = \frac{1}{4} \tan{\frac{\xi_i}{2} } \left( 1 + e^{\mathrm{i}\xi_1} \right) \left( 1 + e^{\mathrm{i}\xi_2} \right) \left( 1 + e^{\mathrm{i}\xi_3} \right)
\end{equation}
where each component of $\boldsymbol{\xi}$ is equispaced following eq. \eqref{freqs}.

\section{Calculation of the dislocation Nye tensor}\label{App1}

The methodology to compute the dislocation Nye tensor $\mathbf{G}$ using the properties of the Fourier transform is presented in this appendix. According to the geometrical considerations presented by \cite{Nye1953-1}, the relationship between the crystallographic slip and the Burgers vector $\mathbf{b}$, can be determined from the closure failure of a Burgers circuit $c$ on surface $S$ with plane normal $N$. Under these considerations, and using Stokes theorem, the expression reads as
\begin{equation}\label{eq:bur}
\mathbf{b}=\oint_{c_0}\mathbf{F}^p\left(\mathbf{x}\right) \, \rd\mathbf{X} =-\int_{S_0}\nabla_0 \times\mathbf{F}^p\left(\mathbf{x}\right) \cdot \mathbf{N}\, \rd\mathbf{S} =\int_{S_0}\boldsymbol{\alpha}\left( \mathbf{x} \right) \cdot \mathbf{N} \, \rd\mathbf{S} 
\end{equation}

\noindent and, thus, the definition of the dislocation density (Nye's) tensor corresponds to

\begin{equation}\label{eq:nyeindex}
G_{ij}=-\nabla \times  \mathbf{F}^p.
\end{equation}

This linear differential operation can be evaluated at a time for all the points in the RVE by computing the curl operation in the Fourier space,  using the discrete derivation rules given in Appendix A. The expression of this operation reads

\begin{equation}\label{eq:nyeindexfourier}
 G_{ij}=\mathcal{F}^{-1}\{ \epsilon_{mli}  \mathcal{F}(\mathbf{F}^p_{jl}) k_m\  \} \text{,}
\end{equation}

\noindent where $\mathcal{F}$ and $\mathcal{F}^{-1}$ stand for the Fourier and inverse Fourier transform of a field, $k_m$  corresponds to the component $m$ of the modified frequency vector, eq. \eqref{kas}, and  $\boldsymbol{\epsilon}$ is the Levi-Civita permutation tensor. In this definition, the use of the discrete differential operator is crucial since the plastic deformation gradient field may be discontinuous at the grain boundary, eq. \eqref{eq:nyeindexfourier}, and the standard derivation leads to the development of Gibbs oscillations, that result in unrealistic values of the Nye tensor. It was observed that this strategy alleviates the Gibbs oscillations and provides accurate values for the Nye tensor. 

\begin{figure}[ht]
\includegraphics[width=0.49\textwidth]{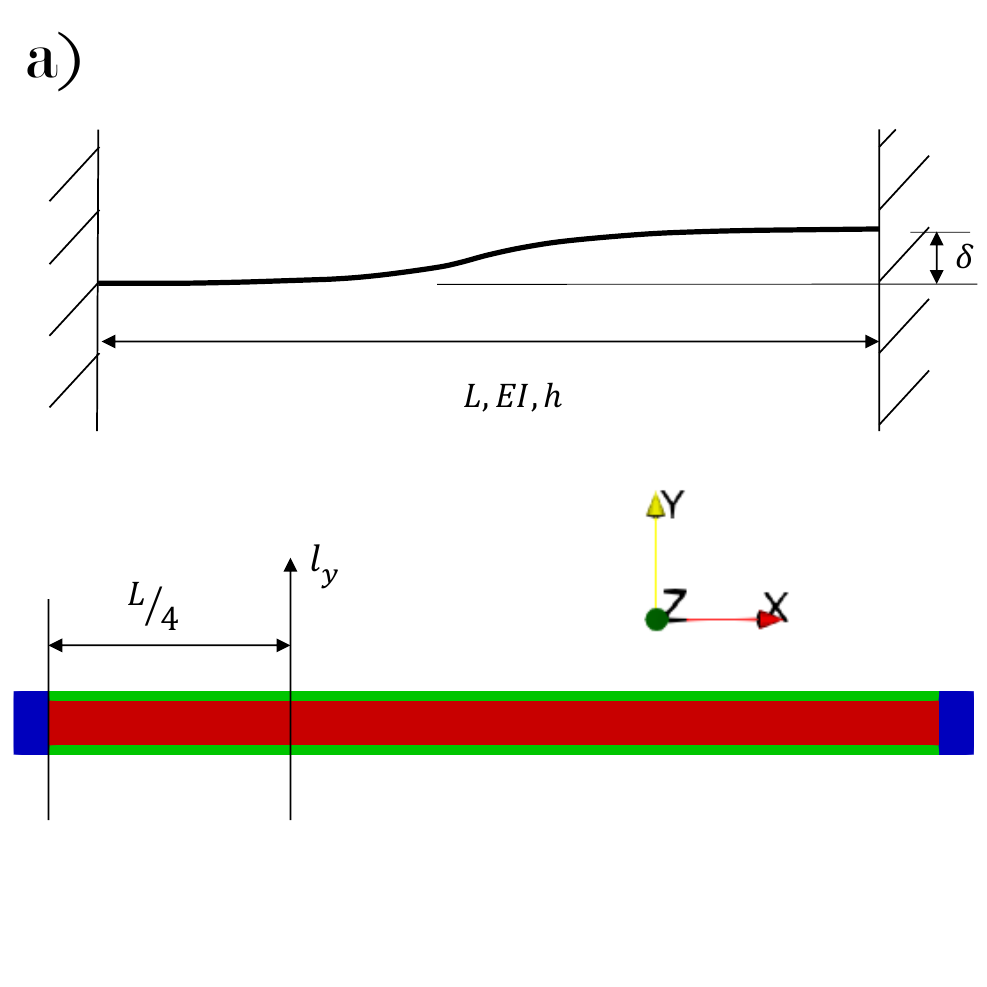}
\includegraphics[width=0.49\textwidth]{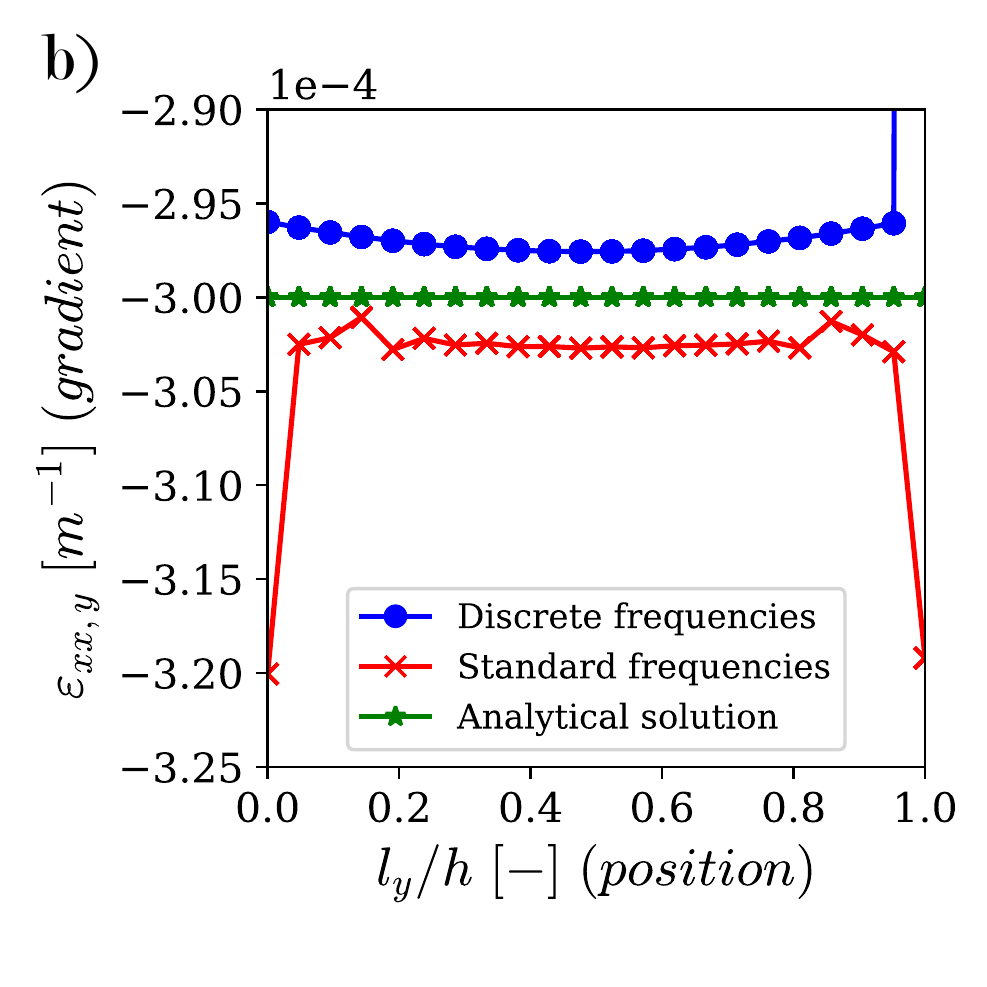}
\caption{(a) Geometry of the encastred beam subjected to simple shear. (b) Elastic strain gradient $\epsilon_{xx,y}$ as a function of the beam depth.}
\label{fig:beam}
\end{figure}

A simple example of application of this strategy is presented below to  benchmark the use of the discrete operator.  An elastic beam  of  length $L$ = 10m  and depth $h$ = $L/20$ is fixed at both ends and subjected to simple shear by displacing one end by an amount $\delta$ = $L$/1000 (Fig. \ref{fig:beam}a). This problem is solved numerically using the FFT in a periodic cell containing three phases: a stiff phase in blue (that fixes the beam), a soft phase in green in the upper and lower sections of the beam (to simulate free boundaries) and the beam in red. Simulations are carried out under plane stress conditions assuming that the behavior of the beam is linear elastic. Under these conditions, the behavior of the beam can be obtained analytically using the Strength of Materials theory.

\begin{figure}[ht]
\includegraphics[width=0.99\textwidth]{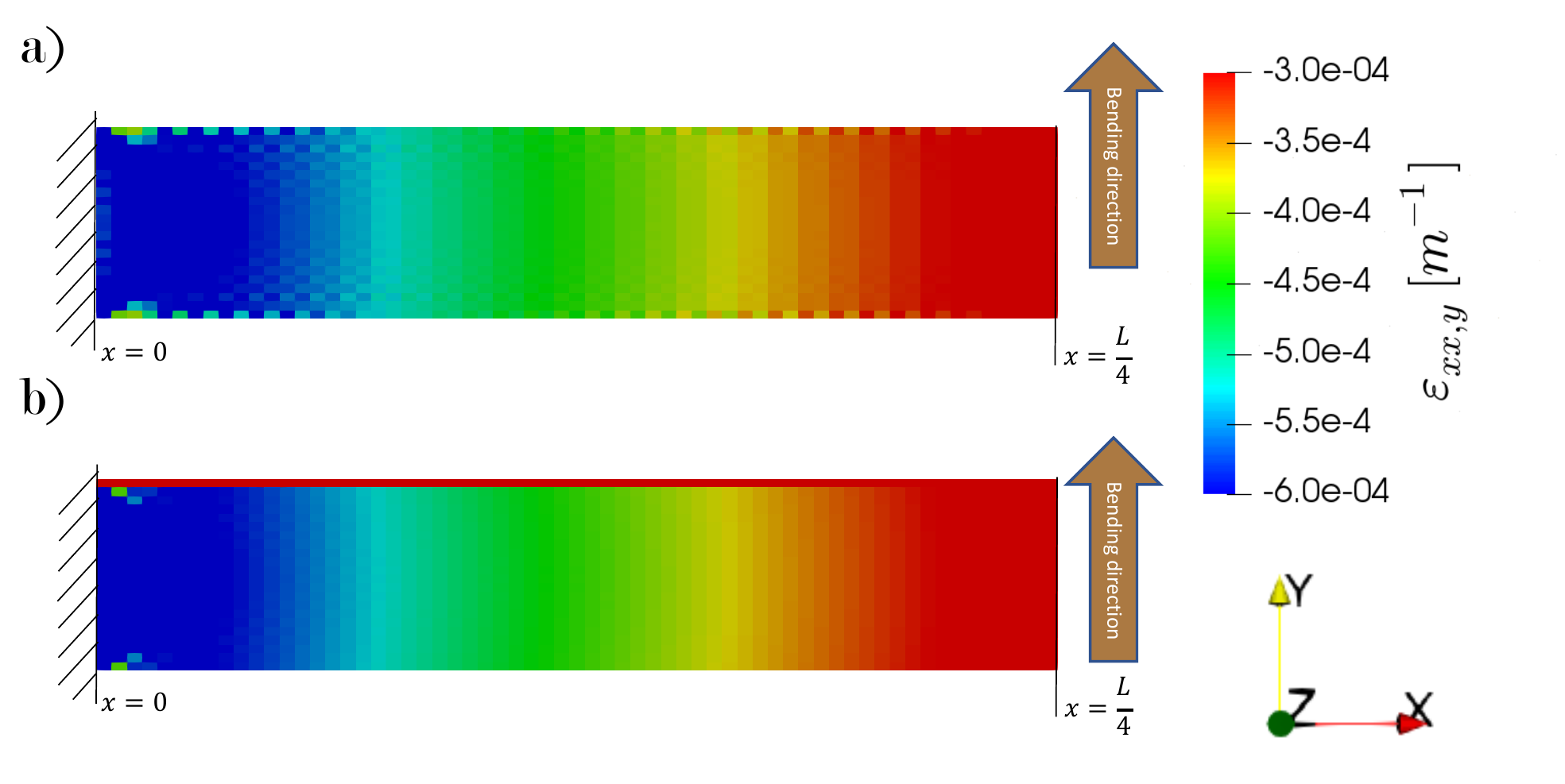}
\caption{Contour plot of the gradient of the strain field $\epsilon_{xx,y}$ for the beam section $x\in [0,L/4]$ obtained by FFT. (a) Standard frequencies. (b) Discrete frequencies.}
\label{fig:grad}
\end{figure}

The macroscopic results (forces in the encastred ends and beam shape) obtained analytically are identical to the  FFT results. However,  differences in the strain gradients are found between the solution obtained using standard frequencies and discrete frequencies in the FFT formulation. The gradient of the elastic deformation in the direction perpendicular to the beam axis, $\epsilon_{xx,y}$, is plotted along the beam depth in Fig. \ref{fig:grad}b. Although the differences in the average curvature with respect to the  analytical solution is $< $ 3\% for both FFT approaches,  the standard FFT simulation presents strong oscillations near the interface with the soft material on the upper and lower sections of the beam. On the contrary, the solution obtained with the modified frequencies shows a smooth profile, indicating that the Gibbs oscillations have been alleviated. The values in the voxels near the boundary differ from the analytical solution, a typical result obtained when a finite-difference derivation is used in a discontinuity. 

The contour plot of the gradient field, $\epsilon_{xx,y}$, obtained using both FTT approaches is plotted in Fig. \ref{fig:grad} for the region of the beam $x\in [0,L/4]$. The Gibbs oscillations (characterized by the chessboard pattern)  are observed when using the standard frequency version, Fig. \ref{fig:grad}a. On the contrary, the contour plot of the gradient obtained using discrete derivatives is smooth in the interior of the beam and only presents unrealistic values in the voxel layer of the upper surface, which corresponds to the soft material in the boundary.

\bibliographystyle{elsarticle-harv}

\section*{References}
\begin{thebibliography}{76}
\expandafter\ifx\csname natexlab\endcsname\relax\def\natexlab#1{#1}\fi
\expandafter\ifx\csname url\endcsname\relax
  \def\url#1{\texttt{#1}}\fi
\expandafter\ifx\csname urlprefix\endcsname\relax\def\urlprefix{URL }\fi

\bibitem[{Acharya and Bassani(1995)}]{ACHARYA1995}
Acharya, A., Bassani, J., 1995. Incompatible lattice deformations and crystal
  plasticity. Plastic and Fracture Instabilities in Materials, ASME proceedings
  57, 75--80.

\bibitem[{Acharya et~al.(2003)Acharya, Bassani, and Beaudoin}]{Acharya2003-1}
Acharya, A., Bassani, J.~L., Beaudoin, A., 2003. Geometrically necessary
  dislocations, hardening, and a simple gradient theory of crystal plasticity.
  Scripta Materialia 48, 167--172.

\bibitem[{Acharya and Beaudoin(2000)}]{AB00}
Acharya, A., Beaudoin, A.~J., 2000. Grain size effects in viscoplastic
  polycrystals at moderate strains. Journal of the Mechanics and Physics of
  Solids 48, 2213--2230.

\bibitem[{Aifantis(1987)}]{AIFANTIS1987211}
Aifantis, E.~C., 1987. The physics of plastic deformation. International
  Journal of Plasticity 3, 211 -- 247.

\bibitem[{Alankar et~al.(2009)Alankar, Mastorakos, and
  Field}]{Al-elastic-constants}
Alankar, A., Mastorakos, I.~N., Field, D.~P., 2009. {A
  dislocation-density-based 3D crystal plasticity model for pure aluminum}.
  Acta Materialia 57, 5936--5946.

\bibitem[{Armstrong et~al.(1962)Armstrong, Codd, Douthwaite, and Petch}]{ACD62}
Armstrong, R., Codd, I., Douthwaite, R.~M., Petch, N.~J., 1962. The plastic
  deformation of polycrystalline aggregates. Philosophical Magazine 7, 45--58.

\bibitem[{Arsenlis and Parks(1999)}]{ARSENLIS19991597}
Arsenlis, A., Parks, D., 1999. Crystallographic aspects of
  geometrically-necessary and statistically-stored dislocation density. Acta
  Materialia 47, 1597 -- 1611.

\bibitem[{Ashby(1970)}]{A70}
Ashby, M., 1970. The deformation of plastically non-homogeneous materials.
  {P}hilosophical {M}agazine 21, 399--424.

\bibitem[{Balint et~al.(2008)Balint, Deshpande, Needleman, and {Van der
  Giessen}}]{BDN08}
Balint, D.~S., Deshpande, V.~S., Needleman, A., {Van der Giessen}, E., 2008.
  Discrete dislocation plasticity analysis of the grain size dependence of the
  flow strength of polycrystals. International Journal of Plasticity 24,
  2149--2172.

\bibitem[{Bardella(2006)}]{BARDELLA2006128}
Bardella, L., 2006. A deformation theory of strain gradient crystal plasticity
  that accounts for geometrically necessary dislocations. Journal of the
  Mechanics and Physics of Solids 54, 128 -- 160.

\bibitem[{Bardella et~al.(2013)Bardella, Segurado, Panteghini, and
  LLorca}]{BSP13}
Bardella, L., Segurado, J., Panteghini, A., LLorca, J., 2013. Latent hardening
  size effect in small-scale plasticity. Modelling and Simulation in Materials
  Science and Engineering 21, 055009.

\bibitem[{Bargmann et~al.(2010)Bargmann, Ekh, Runesson, and Svendsen}]{BER10}
Bargmann, S., Ekh, M., Runesson, K., Svendsen, B., 2010. Modeling of
  polycrystals with gradient crystal plasticity: A comparison of strategies.
  Philosophical Magazine 90, 1263--1288.

\bibitem[{Bayley et~al.(2007)Bayley, Brekelmans, and Geers}]{BBG07}
Bayley, C.~J., Brekelmans, W. A.~M., Geers, M. G.~D., 2007. A three-dimensional
  dislocation field crystal plasticity approach applied to miniaturized
  structures. Philosophical Magazine 87, 1361--1378.

\bibitem[{Berbenni et~al.(2014)Berbenni, Taupin, Djaka, and
  Fressengeas}]{BERBENNI2014}
Berbenni, S., Taupin, V., Djaka, K.~S., Fressengeas, C., 2014. A numerical
  spectral approach for solving elasto-static field dislocation and
  g-disclination mechanics. International Journal of Solids and Structures 51,
  4157 -- 4175.

\bibitem[{Bieler et~al.(2019)Bieler, Alizadeh, Pe{\~n}a-Ortega, and
  LLorca}]{BAP19}
Bieler, T.~R., Alizadeh, R., Pe{\~n}a-Ortega, M., LLorca, J., 2019. An analysis
  of (the lack of) slip transfer between near-cube oriented grains in pure
  {Al}. International Journal of Plasticity 118, 269--290.

\bibitem[{Busso et~al.(2000)Busso, Meissonnier, and O'Dowd}]{Busso2000-1}
Busso, E.~P., Meissonnier, F.~T., O'Dowd, N.~P., 2000. Gradient-dependent
  deformation of two-phase single crystals. Journal of the Mechanics and
  Physics of Solids 48, 2333--2361.

\bibitem[{Carreker(1957)}]{ag-exp-dat}
Carreker, R.~P., 1957. {Tensile deformation of silver as a function of
  temperature, strain rate, and grain size}. JOM 9, 112--115.

\bibitem[{Castelluccio and McDowell(2017)}]{Ni-elastic-constants}
Castelluccio, G.~M., McDowell, D.~L., 2017. {Mesoscale cyclic crystal
  plasticity with dislocation substructures}. International Journal of
  Plasticity 57, 1--26.

\bibitem[{Cermelli and Gurtin(2001)}]{CERMELLI20011539}
Cermelli, P., Gurtin, M.~E., 2001. On the characterization of geometrically
  necessary dislocations in finite plasticity. Journal of the Mechanics and
  Physics of Solids 49, 1539 -- 1568.

\bibitem[{Cheong et~al.(2005)Cheong, Busso, and Arsenlis}]{CBA05}
Cheong, K.~S., Busso, E.~P., Arsenlis, A., 2005. A study of microstructural
  length scale effects on the behavior of {FCC} polycrystals using strain
  gradient concepts. International {J}ournal of {P}lasticity 21, 1797--1814.

\bibitem[{Dai and Parks(1997)}]{Dai1997-1}
Dai, H., Parks, D.~M., 1997. Geometrically-necessary dislocation density and
  scale-dependent crystal plasticity. In: Khan, A. (Ed.), Proceedings of Sixth
  International Symposium on Plasticity. Gordon \&\ Breach, pp. 17--18.

\bibitem[{de~Geus et~al.(2017)de~Geus, Vondrejc, Zeman, Peerlings, and
  Geers}]{Geers2016}
de~Geus, T., Vondrejc, J., Zeman, J., Peerlings, R., Geers, M., 2017. Finite
  strain fft-based non-linear solvers made simple. Computer Methods in Applied
  Mechanics and Engineering 318, 412 -- 430.

\bibitem[{Delaire et~al.(2000)Delaire, Raphanel, and Rey}]{DRC00}
Delaire, F., Raphanel, J.~L., Rey, C., 2000. Plastic heterogeneities of a
  copper multicrystal deformed in uniaxial tension: experimental study and
  finite element simulations. Acta Materialia 48, 1075 -- 1087.

\bibitem[{Devincre and Kubin(2010)}]{interaction-matrix-coefficients}
Devincre, B., Kubin, L., 2010. {Scale transitions in crystal plasticity by
  dislocation dynamics simulations}. C. R. Physique 11, 274--284.

\bibitem[{{Di Leo} and Rimoli(2019)}]{DR19}
{Di Leo}, C.~V., Rimoli, J.~J., 2019. New perspectives on the grain-size
  dependent yield strength of polycrystalline metals. Scripta Materialia 166,
  149 -- 153.

\bibitem[{Djaka et~al.(2019)Djaka, Berbenni, Taupin, and Lebensohn}]{DJAKA2019}
Djaka, K.~S., Berbenni, S., Taupin, V., Lebensohn, R.~A., 2019. A {FFT}-based
  numerical implementation of mesoscale field dislocation mechanics:
  Application to two-phase laminates. International Journal of Solids and
  Structures, in press.

\bibitem[{DREAM.3D(2016)}]{DREAM3D}
DREAM.3D, 2016. http://www.dream3d.bluequartz.net.

\bibitem[{Dunne et~al.(2007)Dunne, Rugg, and Walker}]{DUNNE20071061}
Dunne, F., Rugg, D., Walker, A., 2007. Length scale-dependent, elastically
  anisotropic, physically-based hcp crystal plasticity: Application to
  cold-dwell fatigue in {Ti} alloys. International Journal of Plasticity 23,
  1061 -- 1083.

\bibitem[{Dunstan and Bushby(2013)}]{DB13}
Dunstan, D.~J., Bushby, A.~J., 2013. The scaling exponent in the size effect of
  small scale plastic deformation. International Journal of Plasticity 40,
  152--162.

\bibitem[{Dunstan and Bushby(2014)}]{DB14}
Dunstan, D.~J., Bushby, A.~J., 2014. Gran size dependence of the strength of
  metals: The {Hall-Petch} effect does not scale as the inverse of the square
  root of the grain size. International Journal of Plasticity 53, 55--65.

\bibitem[{Eisenlohr et~al.(2013)Eisenlohr, Diehl, Lebensohn, and
  Roters}]{Eisenlohr2013}
Eisenlohr, P., Diehl, M., Lebensohn, R.~A., Roters, F., 2013. A spectral method
  solution to crystal elasto-viscoplasticity at finite strains. International
  Journal of Plasticity 46, 37--53.

\bibitem[{El-Awady(2015)}]{E15}
El-Awady, J.~A., 2015. Unraveling the physics of size-dependent
  dislocation-mediated plasticity. Nature Communications 6, 5926.

\bibitem[{Franciosi et~al.(1980)Franciosi, Berveiller, and Zaoui}]{FBZ80}
Franciosi, P., Berveiller, M., Zaoui, A., 1980. Latent hardening in copper and
  aluminium single crystals. Acta Metallurgica 28, 273--283.

\bibitem[{Fu et~al.(2001)Fu, Benson, and Meyers}]{FBM01}
Fu, H.-H., Benson, D.~J., Meyers, M.~A., 2001. Analytical and computational
  description of effect of grain size on yield stress of metals. Acta
  Materialia 49, 2567--2582.

\bibitem[{Gurtin(2002)}]{GURTIN20025}
Gurtin, M.~E., 2002. A gradient theory of single-crystal viscoplasticity that
  accounts for geometrically necessary dislocations. Journal of the Mechanics
  and Physics of Solids 50, 5 -- 32.

\bibitem[{Gurtin(2008{\natexlab{a}})}]{GURTIN2008702}
Gurtin, M.~E., 2008{\natexlab{a}}. A finite-deformation, gradient theory of
  single-crystal plasticity with free energy dependent on densities of
  geometrically necessary dislocations. International Journal of Plasticity 24,
  702 -- 725.

\bibitem[{Gurtin(2008{\natexlab{b}})}]{Gurtin2008}
Gurtin, M.~E., 2008{\natexlab{b}}. A theory of grain boundaries that accounts
  automatically for grain misorientation and grain-boundary orientation.
  Journal of the Mechanics and Physics of Solids 56, 640 -- 662.

\bibitem[{Gurtin and Needleman(2005)}]{Gurtin2005}
Gurtin, M.~E., Needleman, A., 2005. Boundary conditions in small-deformation,
  single-crystal plasticity that account for the burgers vector. Journal of the
  Mechanics and Physics of Solids 53, 1 -- 31.

\bibitem[{Hall(1951)}]{Hall1951-1}
Hall, E.~O., 1951. The deformation and ageing of mild steel: {III} {D}iscussion
  of results. Proceedings of the Physical Society Section B 64, 747--753.

\bibitem[{Han et~al.(2005{\natexlab{a}})Han, Gao, Huang, and Nix}]{HAN20051188}
Han, C.-S., Gao, H., Huang, Y., Nix, W.~D., 2005{\natexlab{a}}. Mechanism-based
  strain gradient crystal plasticity €"i. theory. Journal of the Mechanics and
  Physics of Solids 53, 1188 -- 1203.

\bibitem[{Han et~al.(2005{\natexlab{b}})Han, Gao, Huang, and Nix}]{HAN20051204}
Han, C.-S., Gao, H., Huang, Y., Nix, W.~D., 2005{\natexlab{b}}. Mechanism-based
  strain gradient crystal plasticity €"ii. analysis. Journal of the Mechanics
  and Physics of Solids 53, 1204 -- 1222.

\bibitem[{Hansen(1977)}]{al-exp-dat}
Hansen, N., 1977. {The effect of grain size and strain on the tensile flow
  stress of aluminium at room temperature}. Acta Metallurgica 25, 863--869.

\bibitem[{Hansen and Ralph(1982)}]{HR82}
Hansen, N., Ralph, B., 1982. The strain and grain size dependence of the flow
  stress of copper. Acta Metallurgica 30, 411 -- 417.

\bibitem[{Haouala et~al.(2018)Haouala, Segurado, and LLorca}]{HSL18}
Haouala, S., Segurado, J., LLorca, J., 2018. {An analysis of the influence of
  grain size on the strengthof FCC polycrystals by means of computational
  homogenization}. Acta Materialia 148, 72--85.

\bibitem[{H\'emery et~al.(2018)H\'emery, Nizou, and Villechaise}]{HNV18}
H\'emery, S., Nizou, P., Villechaise, P., 2018. In situ sem investigation of
  slip transfer in ti-6al-4v: Effect of applied stress. Materials Science and
  Engineering A 709, 277--284.

\bibitem[{Hirth(1972)}]{H72}
Hirth, J.~P., 1972. The influence of grain boundaries on mechanical properties.
  Metallurgical Transactions 3, 3047--3067.

\bibitem[{Kocks(1970)}]{K70}
Kocks, U., 1970. The relation between polycrystal deformation and single
  crystal deformation. Metallurgical {T}ransactions 1, 1121--1143.

\bibitem[{Kocks and Mecking(1981)}]{KM81}
Kocks, U.~F., Mecking, H., 1981. Kinetics of flow and strain-hardening. Acta
  Metallurgica 29, 1865--1875.

\bibitem[{Lebensohn et~al.(2004)Lebensohn, Liu, and {Ponte
  Casta{\~n}eda}}]{LLP04}
Lebensohn, R., Liu, Y., {Ponte Casta{\~n}eda}, P., 2004. On the accuracy of the
  self-consistent approximation for polycrystals: comparison with full-field
  numerical simulations. Acta Materialia 52, 5347 -- 5361.

\bibitem[{Lebensohn(2001)}]{L01}
Lebensohn, R.~A., 2001. N-site modelling of a 3d viscoplastic polycrystal using
  {Fast Fourier Transform}. Acta Materialia 49, 2723--2737.

\bibitem[{Lebensohn and Needleman(2016)}]{LN16}
Lebensohn, R.~A., Needleman, A., 2016. Numerical implementation of non-local
  polycrystal plasticity using {Fast Fourier Transforms}. Journal of the
  Mechanics and Physics of Solids 97, 333--351.

\bibitem[{Li et~al.(2016)Li, Bushby, and Dunstan}]{LBD16}
Li, Y., Bushby, A.~J., Dunstan, D.~J., 2016. The {Hall-Petch} effect as a
  manifestation of a general size effect. Proceedings of the Royal Society A
  472, 20150890.

\bibitem[{Lucarini and Segurado(2019{\natexlab{a}})}]{LS2019b}
Lucarini, S., Segurado, J., 2019{\natexlab{a}}. An algorithm for stress and
  mixed control in galerkin-based fft homogenization. International Journal for
  Numerical Methods in Engineering 119, 797--805.

\bibitem[{Lucarini and Segurado(2019{\natexlab{b}})}]{LS2019a}
Lucarini, S., Segurado, J., 2019{\natexlab{b}}. On the accuracy of spectral
  solvers for micromechanics based fatigue modeling. Computational Mechanics
  63, 365--382.

\bibitem[{Ma et~al.(2006)Ma, Roters, and Raabe}]{Ma20062169}
Ma, A., Roters, F., Raabe, D., 2006. A dislocation density based constitutive
  model for crystal plasticity fem including geometrically necessary
  dislocations. Acta Materialia 54, 2169--2179.

\bibitem[{Martinez-Mardones et~al.(2000)Martinez-Mardones, Walgraef, and
  W\"orner}]{MWW00}
Martinez-Mardones, J., Walgraef, D., W\"orner, C., 2000. Materials
  Instabilities. World Scientific.

\bibitem[{Mohazzabi(1985)}]{Ag-elastic-constants}
Mohazzabi, P., 1985. {Temperature Dependence of the Elastic Constants of
  Copper, Gold and Silver}. Journal of Physics and Chemistry of Solids 46,
  147--150.

\bibitem[{Moulinec and Suquet(1998)}]{MS98}
Moulinec, H., Suquet, P., 1998. A numerical method for computing the overall
  response of nonlinear composites with complex microstructure. Computational
  Methods in Applied Mechanics and Engineering 157, 69--94.

\bibitem[{M{\"u}ller(1998)}]{Mueller98}
M{\"u}ller, W.~H., 1998. Fourier transforms and their application to the
  formation of textures and changes of morphology in solids. In: Bahei-El-Din,
  Y.~A., Dvorak, G.~J. (Eds.), IUTAM Symposium on Transformation Problems in
  Composite and Active Materials. Kluwer, pp. 61--72.

\bibitem[{Narutani and Takamura(1991)}]{ni-exp-dat}
Narutani, T., Takamura, J., 1991. {Grain-size strengthening in terms of
  dislocation density measured by resistivity}. Acta Metallurgica et Materialia
  39, 2037--2049.

\bibitem[{Niordson and Kysar(2014)}]{NIORDSON201431}
Niordson, C.~F., Kysar, J.~W., 2014. Computational strain gradient crystal
  plasticity. Journal of the Mechanics and Physics of Solids 62, 31 -- 47.

\bibitem[{Nye(1953)}]{Nye1953-1}
Nye, J.~F., 1953. Some geometrical relations in dislocated crystals. Acta
  metallurgica 1, 153--162.

\bibitem[{Petch(1953)}]{Petch1953-1}
Petch, N.~J., 1953. The cleavage strength of polycrystals. Journal of the Iron
  and Steel Institute 174, 25--28.

\bibitem[{Raj and Pharr(1986)}]{RP86}
Raj, S.~V., Pharr, G.~M., 1986. A compilation and analysis for the stress
  dependence of the subgrain size. Materials Science and Engineering A 81,
  217--237.

\bibitem[{Roters et~al.(2019)Roters, Diehl, Shanthraj, Eisenlohr, Reuber, Wong,
  Maiti, Ebrahimi, Hochrainer, Fabritius, Nikolov, Friák, Fujita, Grilli,
  Janssens, Jia, Kok, Ma, Meier, Werner, Stricker, Weygand, and
  Raabe}]{ROTERS2019420}
Roters, F., Diehl, M., Shanthraj, P., Eisenlohr, P., Reuber, C., Wong, S.,
  Maiti, T., Ebrahimi, A., Hochrainer, T., Fabritius, H.-O., Nikolov, S.,
  Friák, M., Fujita, N., Grilli, N., Janssens, K., Jia, N., Kok, P., Ma, D.,
  Meier, F., Werner, E., Stricker, M., Weygand, D., Raabe, D., 2019. Damask:
  The dusseldorf advanced material simulation kit for modeling multi-physics
  crystal plasticity, thermal, and damage phenomena from the single crystal up
  to the component scale. Computational Materials Science 158, 420 -- 478.

\bibitem[{Roters et~al.(2010)Roters, Eisenlohr, Hantcherli, Tjahjanto, Bieler,
  and Raabe}]{REH10}
Roters, F., Eisenlohr, P., Hantcherli, L., Tjahjanto, D.~D., Bieler, T.~R.,
  Raabe, D., 2010. Overview of constitutive laws, kinematics, homogenization
  and multiscale methods in crystal plasticity finite-element modeling: theory,
  experiments, applications. Acta Materialia 58, 1152--1211.

\bibitem[{Rovinelli et~al.(2019)Rovinelli, Proudhon, Lebensohn, and
  Sangid}]{ROVINELLI2019}
Rovinelli, A., Proudhon, H., Lebensohn, R.~A., Sangid, M.~D., 2019. Assessing
  the reliability of fast fourier transform-based crystal plasticity
  simulations of a polycrystalline material near a crack tip. International
  Journal of Solids and Structures, in press.

\bibitem[{Rubio et~al.(2019)Rubio, Haouala, and LLorca}]{RHL19}
Rubio, R.~A., Haouala, S., LLorca, J., 2019. Grain boundary strengthening of
  {FCC} polycrystals. Journal of Materials Research 34, 2263 -- 2274.

\bibitem[{Sauzay and Kubin(2011)}]{SK11}
Sauzay, M., Kubin, L., 2011. Scaling laws for dislocation microstructures in
  monotonic and cyclic deformation of fcc metals. Progress in Materials Science
  56, 725 -- 784.

\bibitem[{Segurado et~al.(2018)Segurado, Lebensohn, and LLorca}]{SLL18}
Segurado, J., Lebensohn, R.~A., LLorca, J., 2018. Computational homogenization
  of polycrystal. Advances in Applied Mechanics 51, 1 -- 114.

\bibitem[{Shu and Fleck(1999)}]{SHU1999297}
Shu, J.~Y., Fleck, N.~A., 1999. Strain gradient crystal plasticity:
  size-dependent deformation of bicrystals. Journal of the Mechanics and
  Physics of Solids 47, 297 -- 324.

\bibitem[{Vondrejc et~al.(2014)Vondrejc, Zeman, and Marek}]{Vondrejc2014}
Vondrejc, J., Zeman, J., Marek, I., 2014. An fft-based galerkin method for
  homogenization of periodic media. Computers and Mathematics with Applications
  68, 156 -- 173.

\bibitem[{Willot(2015)}]{W15}
Willot, F., 2015. Fourier-based schemes for computing the mechanical response
  of composites with accurate local fields. Comptes Rendus M\'{e}canique 34,
  232--245.

\bibitem[{Zaefferer et~al.(2003)Zaefferer, Kuo, Zhao, Winning, and
  Raabe}]{vicoplastic-parameters-fcc}
Zaefferer, S., Kuo, J.-C., Zhao, Z., Winning, M., Raabe, D., 2003. {On the
  influence of the grain boundary misorientation on the plastic deformation of
  aluminum bicrystals}. Acta Materialia 51, 4719--4735.

\bibitem[{Zaiser and Sandfeld(2014)}]{ZS14}
Zaiser, M., Sandfeld, S., 2014. Scaling properties of dislocation simulations
  in the similitude regime. Modeling and Simulations in Materials Science and
  Engineering 22, 065012.

\bibitem[{Zeman et~al.(2017)Zeman, de~Geus, Vondrejc, Peerlings, and
  Geers}]{Geers2017}
Zeman, J., de~Geus, T. W.~J., Vondrejc, J., Peerlings, R. H.~J., Geers, M.
  G.~D., 2017. A finite element perspective on nonlinear fft-based
  micromechanical simulations. International Journal for Numerical Methods in
  Engineering 110, 903--926.

\end{thebibliography}

\end{document}